\documentclass[reprint,aps,prd,twocolumn,superscriptaddress,nofootinbib,amsmath,amssymb]{revtex4-2}

\usepackage{amsmath,amssymb}
\usepackage{graphicx}
\usepackage{dcolumn}
\usepackage{bm}
\usepackage{color}
\usepackage{booktabs}
\usepackage{tabularx}

\usepackage{orcidlink}
\usepackage{hyperref}
\hypersetup{colorlinks=true,linkcolor=blue,filecolor=blue,urlcolor=blue,citecolor=blue}

\graphicspath{{figures/}}

\newcommand{\Msun}{\ensuremath{M_{\odot}}}

\newcommand{\Neff}{N_{\mathrm{eff}}}

\begin{document}

\preprint{APS/123-QED}

\title{Joint Inference of Strongly Lensed Gravitational Waves and Electromagnetic Strong-Lensing Systems}

\author{Qiyuan Yang \orcidlink{0009-0007-6444-2026}}
\affiliation{School of Physics and Technology, Wuhan University, Wuhan 430072, China}

\author{Hui Li \orcidlink{0009-0005-9594-8375}}
\affiliation{Institute for Astrophysics, School of Physics, Zhengzhou University, Zhengzhou, 450001, People’s Republic of China}

\author{Xiaoyue Cao \orcidlink{0000-0003-4988-9296}}
\affiliation{Institute for Astrophysics, School of Physics, Zhengzhou University, Zhengzhou, 450001, People’s Republic of China}

\author{Rui Li \orcidlink{0000-0002-3490-4089}}
\affiliation{Institute for Astrophysics, School of Physics, Zhengzhou University, Zhengzhou, 450001, People’s Republic of China}
    
\author{Xilong Fan \orcidlink{0000-0002-8174-0128}}
\email{xilong.fan@whu.edu.cn}
\affiliation{School of Physics and Technology, Wuhan University, Wuhan 430072, China}

\date{\today}

\begin{abstract}
Multimessenger observations of strongly lensed gravitational waves (GWs) and their associated electromagnetic (EM) strong-lensing systems provide complementary constraints on GW sources and foreground lenses. We develop a Bayesian framework for combining GW and EM information through two-way posterior reweighting without refitting the imaging data. Using ten strong-lensing systems observed by the Hubble Space Telescope, we simulate pairs of lensed binary-black-hole signals and jointly estimate their parameters using EM-derived source information. Across the ten systems, further incorporating EM lens-model posteriors through reweighting reduces the central 90\% credible-interval widths of the two image magnifications by up to 95.3\% and 68.0\%, respectively, and that of the viewing angle by up to 52.2\%. Conversely, GW magnification measurements yield maximum reductions of 62.3\% and 54.0\% in the interval widths of the normalized source--lens offset and mass--light centroid offset, respectively. The corresponding maximum reductions for the Einstein radius, ellipticity amplitude, and external-shear amplitude are 10.8\%, 40.5\%, and 21.5\%, respectively. GW time-delay measurements provide additional constraints, but direct reweighting causes severe weight concentration. Introducing time-delay modeling uncertainty reduces this concentration at the cost of weaker timing constraints. Our results demonstrate the potential of joint GW--EM analysis for constraining GW source properties, source--lens geometry, and the relative alignment of the mass and light distributions of foreground lenses.
\end{abstract}

\maketitle

\section{Introduction}\label{sec:introduction}
The observation of gravitational waves (GWs) from a binary-black-hole merger opened a direct window on compact-binary dynamics~\citep{Abbott2016GW150914}. Observations with Advanced LIGO and Advanced Virgo~\citep{AdvancedLIGO,AdvancedVirgo} have established growing catalogs of compact-binary mergers~\citep{GWTC1,GWTC2,GWTC3}, whose waveforms encode masses, spins, orientation, and luminosity distance~\citep{SathyaprakashSchutz2009}. Future observatories, including the Einstein Telescope (ET) and Cosmic Explorer, will extend these measurements to greater cosmological distances~\citep{Maggiore2020ET,CosmicExplorer}, increasing the opportunities to study sources affected by gravitational lensing. Gravitational lensing connects distant sources to intervening matter through a common framework for GW and electromagnetic (EM) observations~\citep{NarayanBartelmann1996,Wambsganss1998,Bartelmann2010}. A foreground galaxy can produce multiple images of one merger, with different amplitudes and arrival times~\citep{Oguri2019Transients}. Predictions for detectable lensed-GW populations depend on source and lens distributions, magnification bias, and detector sensitivity~\citep{Ng2018Rates,Li2018Statistics,Oguri2018,Wierda2021}. For a smooth galaxy-scale lens at ground-based GW frequencies, geometric optics generally applies: each image preserves the intrinsic source evolution but acquires a magnification, arrival-time shift, and image-type-dependent phase~\citep{TakahashiNakamura2003,DaiVenumadhav2017,Ezquiaga2021}. The measured amplitude scales as $\sqrt{|\mu|}/d_L$ and also depends on binary orientation and detector response. Unknown magnification can therefore bias the inferred source distance and masses~\citep{Dai2017Magnification}, while correlations among amplitude-related parameters complicate its measurement~\citep{CutlerFlanagan1994,HolzHughes2005}. These degeneracies motivate the incorporation of independent EM information to improve GW parameter estimation. More generally, complementary EM observations can provide additional constraints on GW source properties, following the broader strategy of multimessenger astronomy. For example, \citet{Fan2014Multimessenger} developed a Bayesian framework incorporating host-galaxy information to improve source localization and inclination estimates, while \citet{Fan2017GRB} demonstrated the feasibility of jointly constraining binary parameters and short gamma-ray burst (SGRB) luminosities.

For lensed GW events associated with identified EM lens systems, the relevant EM information can be obtained from the foreground lens galaxy and the background host galaxy, without necessarily requiring a transient EM counterpart associated with the binary merger~\citep{Yu2020Localization}. Spectroscopic observations provide the redshifts of both the foreground lens and the background source, while high-resolution imaging constrains the lens mass distribution and the structure of the background source. The Sloan Lens ACS Survey (SLACS) provides real lens systems with Hubble Space Telescope (HST) imaging and spectroscopic observations suitable for this purpose~\citep{Bolton2006,REF_Bolton2008}. In particular, once the source redshift is known, the luminosity distance can be determined under an assumed cosmological model, helping to reduce the degeneracy between luminosity distance and lensing magnification in GW amplitude measurements. Furthermore, at a specified source position, the posterior distribution of the EM lens model provides joint predictions for the absolute magnifications and relative time delays of the multiple images~\citep{NarayanBartelmann1996,Yu2020Localization,Wempe2024Localization}. These predictions provide complementary constraints on the GW signals, while the magnification and time-delay measurements from GWs can, in turn, be used to update the EM lens-model posterior and improve constraints on the physical properties of the foreground lens. This mutual exchange of information provides a basis for joint GW--EM parameter estimation in identified strong-lensing systems.

Although no strongly lensed GW event has been conclusively identified to date, considerable efforts have been devoted to searching for multiple images of lensed GWs in existing detector data. These searches exploit the fact that multiple images of the same merger share consistent intrinsic waveform properties, while differing in their magnifications, arrival times, and phase shifts. Several approaches have been developed for identifying potential lensed image pairs, including posterior-overlap methods that compare independently inferred GW parameter posteriors~\citep{Haris2018LensedGW} and joint parameter estimation methods that analyze multiple candidate signals under a common source model~\citep{Liu2021Joint,Lo2023LensedGW}. Dedicated searches have applied these methods to the observed GW catalogs~\citep{Hannuksela2019Search,LVK2021Lensing,LVK2023Lensing}. For example, GOLUM improves the computational efficiency of joint analyses by using the posterior information from one image to inform the parameter estimation of another~\citep{Janquart2021GOLUM,Janquart2023GOLUM}.

Beyond GW-only searches, multimessenger studies have explored the use of EM information to assist in identifying lensed GW events and their host lens systems. Hannuksela et al.~\citep{Hannuksela2020Localization} demonstrated the potential for host-galaxy identification and source-plane localization through joint GW--EM observations. Wempe et al.~\citep{Wempe2024Localization} investigated these prospects for lens systems observable with a Euclid-like survey, while Liu and Liao~\citep{LiuLiao2026} incorporated galaxy-catalog-based positional constraints into the identification of lensed GW events. In addition to event identification, the reconstruction of lens properties from GW observations has also been investigated. Seo et al.~\citep{Seo2024Halos} studied the constraints on dark-halo properties using different lens models and supplementary EM geometric information, while Poon et al.~\citep{Poon2025Degeneracies} examined the parameter combinations that can be constrained in the presence of similarity-transformation and mass-sheet degeneracies. More recently, \citet{Uronen2026JointBayesian} presented the \texttt{silmarel} framework for joint reconstruction of lensed BBH sources and their host lens systems using GW and EM observations. Collectively, these studies demonstrate the potential of combining GW and EM information for lens identification and reconstruction, while highlighting the importance of accounting for lens-model degeneracies. Although existing approaches have explored how GW information can be incorporated into EM lens reconstruction, a systematic assessment of how information transfer between existing GW and EM posterior distributions improves the constraints on both GW source and lens parameters remains limited.

Motivated by this, we develop a Bayesian framework based on bidirectional posterior reweighting to combine existing GW and EM lens-model posteriors, assuming an identified GW--EM association. Using the source redshift and sky position measured from EM observations, together with an assumed cosmological model, we constrain the absolute magnifications of both GW images and quantify how EM lens-model information improves the measurements of GW magnifications and binary viewing angles. Conversely, we use GW magnification measurements to update the EM lens-model posteriors, examining improvements in the constraints on conventional lens parameters and derived geometrical quantities. A key feature of our approach is that it directly reuses parameter-estimation results independently obtained from GW and EM observations, without requiring repeated evaluations of the original data likelihoods or computationally expensive EM image refitting. This allows information encoded in the original posterior distributions, including parameter correlations, to be incorporated into the joint analysis. We also investigate the impact of time-delay modeling uncertainties and finite posterior sampling on the reliability of the reweighting procedure.

To demonstrate this framework, we apply it to ten real strong-lensing systems observed by the Hubble Space Telescope (HST), using simulated pairs of lensed binary-black-hole signals in an Einstein Telescope (ET) configuration. We jointly analyze the two GW images with shared source parameters and separate magnifications and arrival times, conditioning on the EM sky position and the luminosity distance derived from the source redshift under a fixed cosmological model. We then perform two-way posterior reweighting to quantify the improvements in GW magnification and viewing-angle constraints, as well as the reverse constraints on lens parameters and normalized source--lens offsets. Particular attention is given to the additional information provided by GW relative time-delay measurements and the effects of modeling uncertainties on the stability of the EM posterior update. The ten systems serve as representative case studies for assessing the complementarity of GW and EM observations and the practical applicability of the proposed framework.

This paper is organized as follows. In Sec.~\ref{sec:2}, we describe the modeling of the HST strong-lensing systems, the simulation of lensed GW signals, and the joint inference framework. In Sec.~\ref{sec:3}, we present the resulting improvements in GW and lens-parameter constraints, examine the impact of time-delay modeling uncertainty, and compare the results across the ten systems. Finally, a summary and discussion of this work are provided in Sec.~\ref{sec:4}.
Throughout this paper, we take the speed of light $c=2.998\times 10^8 \mathrm{m}\cdot\mathrm{s}^{-1}$, the gravitational constant $G=6.67\times 10^{-11}\mathrm{m^3}\cdot\mathrm{kg}^{-1}\cdot\mathrm{s^{-2}}$ and the $\Lambda$CDM cosmology model described in Ref.~\cite{Planck2018}, which is also coded in \texttt{Astropy}\footnote{\url{https://www.astropy.org/}}~\citep{Astropy2013,Astropy2018,Astropy2022} with assumptions of  $H_{0}=67.66\mathrm{km}\cdot\mathrm{s}^{-1}\cdot\mathrm{Mpc}^{-1}$ and $\Omega_{\mathrm{m}}=0.30966$.

\begin{figure*}[t]
\centering
\includegraphics[width=\textwidth]{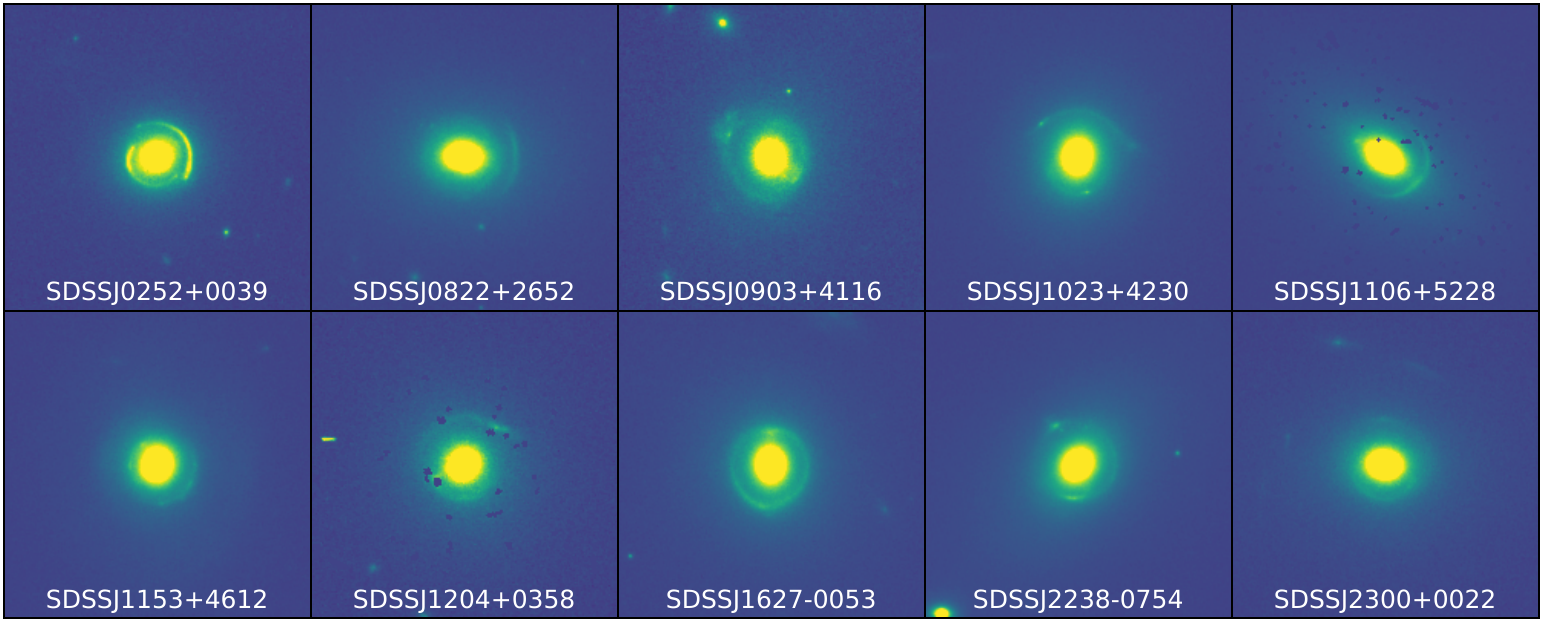}
\caption{The ten HST-observed strong lensing images after the data-reduction procedure. Each system is assumed to host a simulated BBH merger, with the sky location of the GW source fixed to the reference coordinates (RA and DEC) of the corresponding HST strong lensing system.}
\label{fig:hst}
\end{figure*}

\section{Method}\label{sec:2}
\subsection{Strong-lens modeling of HST systems}\label{sec:2.1}

We select ten galaxy-scale strong lensing systems from the Sloan Lens ACS Survey (SLACS; \citealt{REF_Bolton2008}) as illustrative case studies. The systems were chosen because their HST images show prominent arcs or ring-like features from the background source, and have relatively simple lens
configurations that can be described by a singular isothermal
ellipsoid (SIE; \citealt{REF_SIE}) plus external shear. This image-based selection does not by itself establish the number of images predicted for a point source; that is determined from the lens models used in the subsequent analysis. The selected systems do not constitute a statistically complete sample of SLACS. Fig.~\ref{fig:hst} shows their HST images through the \texttt{F814W} filter, and Table~\ref{tab:1} summarizes their coordinates and spectroscopic lens and source redshifts.

We process the HST ACS exposures for these systems following the data-reduction procedure described in \citet{REF_Bolton2008}. Briefly, we calibrate and sky-subtract the exposures, remove cosmic rays, register the images, and resample them bilinearly onto a common grid with a pixel scale of 0.05~arcsec per pixel before combining them. We construct the corresponding noise maps and determine the PSFs for use in the lens modeling. We perform the lens modeling with \texttt{TinyLensGpu}\footnote{\url{https://github.com/caoxiaoyue/TinyLensGpu}}
\citep{REF_TinyLensGpu}, a GPU-based software package for strong lensing modeling. We model the lens mass distribution as an SIE plus external shear. The surface brightness distributions of the lensed source and foreground lens galaxy are represented by multi-Gaussian expansion (MGE) models \citep{REF_MGE}, with ten Gaussian components for the source and fifteen for the lens galaxy. This flexible parameterization allows us to model the foreground lens light and lensed source light separately, helping to reduce lens light contamination in the source reconstruction and to constrain the lens mass model.

The EM-only posterior samples jointly describe the lens mass parameters and reconstructed source position, including their correlations. For each sample, we calculate the point source magnifications and relative time delays, as described in the following subsection. In the subsequent joint GW--EM analysis, we reuse these EM posterior samples and reweight them using the GW data, without refitting the HST images.

\begin{table*}[t]
\centering
\caption{Properties of the ten HST-observed strong-lensing systems adopted for the simulations. The columns list the system name, right ascension (RA) and declination (DEC) of the foreground lens galaxy, lens redshift ($z_l$), source redshift ($z_s$), magnifications ($\mu_0$ and $\mu_1$) and time delay ($\Delta t$) of two images. RA and DEC are given in radians. The magnifications and time delays are derived from the posterior distributions of the EM lens-model parameters, and their median values are presented in the table and used
as the injection parameters for the lensed GW signals.}
\label{tab:1}
\small
\begin{tabular*}{\textwidth}
{@{\extracolsep{\fill}}lcccc|ccc@{}}
\toprule
Name & RA & DEC & $z_l$ & $z_s$
& $\mu_0$ & $\mu_1$ & $\Delta t$ (s) \\
\midrule
J0252+0039 & 0.754 & 0.011 & 0.2803 & 0.9818
& 17.64 & 5.86 & 916374.2487 \\
J0822+2652 & 2.193 & 0.469 & 0.2414 & 0.5941
& 8.47 & 2.00 & 3300110.5329 \\
J0903+4116 & 2.370 & 0.720 & 0.4304 & 1.0645
& 3.43 & 2.82 & 8744901.3222 \\
J1023+4230 & 2.721 & 0.742 & 0.1912 & 0.6960
& 7.93 & 4.02 & 1435801.9701 \\
J1106+5228 & 2.909 & 0.916 & 0.0955 & 0.4069
& 8.40 & 2.10 & 638202.9880 \\
J1153+4612 & 3.112 & 0.806 & 0.1797 & 0.8751
& 1.13 & 0.48 & 129236.9362 \\
J1204+0358 & 3.162 & 0.069 & 0.1644 & 0.6307
& 3.92 & 3.69 & 2058866.1999 \\
J1627$-$0053 & 4.310 & -0.016 & 0.2076 & 0.5241
& 10.83 & 4.50 & 1468899.7986 \\
J2238$-$0754 & 5.928 & -0.138 & 0.1371 & 0.7126
& 3.59 & 5.59 & 1194263.6129 \\
J2300+0022 & 6.025 & 0.007 & 0.2285 & 0.4635
& 2.91 & 4.40 & 3041452.8040 \\
\bottomrule
\end{tabular*}
\end{table*}

\begin{table}[tbp]
\centering
\caption{Parameters of the simulated gravitational-wave signals. The source-frame component masses, spin parameters, orientation angles, polarization angle, coalescence phase, and reference geocentric time of the first image are identical across all ten systems. The geocentric time of the second image is given by $t_1=t_0+\Delta t$, where $\Delta t$ is derived from the EM lens modeling. The luminosity distance is calculated from the source redshift $z_s$, while the sky location is approximated by the catalog coordinates (RA and DEC) of the foreground lens galaxy.}
\label{tab:2}
\small
\begin{tabular*}{\columnwidth}
{@{\extracolsep{\fill}}lc@{}}
\toprule
Parameter & Value \\
\midrule
$m_1^{\rm s}\,(M_\odot)$ & 30.0 \\
$m_2^{\rm s}\,(M_\odot)$ & 20.0 \\
$a_1$ & 0.82 \\
$a_2$ & 0.79 \\
$\theta_1$ (rad) & 2.11 \\
$\theta_2$ (rad) & 1.51 \\
$\phi_{12}$ (rad) & 4.21 \\
$\phi_{JL}$ (rad) & 6.03 \\
$\theta_{JN}$ (rad) & 2.59 \\
$\psi$ (rad) & 2.00 \\
$\phi_c$ (rad) & 5.55 \\
$t_0$ (GPS s) & 1419066219.6282 \\
$t_1$ (GPS s) & $t_0+\Delta t$ \\
$\Delta t$ (s) & System-dependent \\
$d_L$ (Mpc) & System-dependent \\
$\alpha$ (rad) & System-dependent \\
$\delta$ (rad) & System-dependent \\

\bottomrule
\end{tabular*}
\end{table}

\subsection{Simulation of lensed gravitational-wave events}\label{sec:2.2}

\subsubsection{From lens-model samples to image observables}\label{sec:2.2.1}

In the SIE plus external shear model, the parameters relevant to the point-source lens mapping are given by
\begin{equation}\label{eq:1}
\boldsymbol{\lambda}=(x_{\rm L},y_{\rm L},e_1,e_2,\theta_{\rm E},\gamma_1,\gamma_2,\beta_x,\beta_y),
\end{equation}
where $(x_{\rm L},y_{\rm L})$ is the lens mass center, $(e_1,e_2)$ are the ellipticity components, $\theta_{\rm E}$ is the Einstein radius, $(\gamma_1,\gamma_2)$ are the external-shear components, and $(\beta_x,\beta_y)$ is the reconstructed light center of the background source galaxy. Angular coordinates and $\theta_{\rm E}$ are expressed in arcseconds. We evaluate the point-source lens mapping at this reconstructed source-light center to calculate the image magnifications and relative time delay.

For each lens-parameter sample obtained in Sec.~\ref{sec:2.1}, we adopt the fixed lens and source redshifts determined from spectroscopic observations and use \texttt{lenstronomy}\footnote{\url{https://lenstronomy.readthedocs.io/}}~\citep{Lenstronomy2018,Lenstronomy2021} to solve the lens equation and calculate the magnifications and arrival times of the lensed images. For each system, we first determine the image multiplicity using a reference lens model constructed from the component-wise median of the EM posterior parameters. All ten reference models produce two images. For each EM posterior sample, we retain the earliest- and latest-arriving images, which correspond to the two image branches identified in the reference model, and discard any additional solutions. These two images are labeled as image 0 and image 1, respectively. This selection criterion is applied consistently when computing the image magnifications and relative time delays.

We define the derived lensing observables as $\boldsymbol{\gamma}=g(\boldsymbol{\lambda})=(|\mu_0|,|\mu_1|,\Delta t)$, where $\mu_0$ and $\mu_1$ are the signed magnifications of images 0 and 1, respectively, and the relative time delay is defined as $\Delta t=t_1-t_0>0$. The absolute magnifications are adopted to ensure consistency with the parameters inferred from the GW signals.
For each system, we construct the injected lensed GW signals using representative values derived from the EM posterior distributions. Specifically, the injection magnification of each selected image is set to the absolute value of the median of its signed magnification distribution, while the injected relative time delay is taken as the median of the corresponding EM-derived distribution, as shown in Table~\ref{tab:1}. These component-wise medians are adopted as the fiducial injection values and are not necessarily associated with a single realization of the lens model.

\subsubsection{Lensed GW signals}\label{sec:2.2.2}
For each of the ten strong-lensing systems, we simulate a BBH merger hosted by the corresponding background source galaxy. For the lensing calculation, we place the simulated BBH at the reconstructed light center of the background source galaxy, $(\beta_x,\beta_y)$, and solve the lens equation to obtain the image magnifications and relative time delay, as described in Sec.~\ref{sec:2.2.1}. These local source-plane coordinates are distinct from the celestial coordinates used to evaluate the GW detector response. For waveform injection and parameter estimation, we adopt a common sky direction for both lensed images, fixed to the catalog right ascension and declination of the foreground lens galaxy. This approximation neglects the arcsecond-scale offsets of the images from the catalog lens position in the detector-response calculation, while retaining the local source--lens geometry in the calculation of magnifications and lensing time delays. The BBH remains located in the background source galaxy; the foreground lens coordinates serve only as an approximate sky direction for the GW signals.

We adopt identical source-frame component masses $(m_1^{\rm s},m_2^{\rm s})$, spin magnitudes $(a_1,a_2)$, spin orientation angles $(\theta_1,\theta_2,\phi_{12},\phi_{JL})$, viewing angle $\theta_{JN}$, polarization angle $\psi$, coalescence phase $\phi_c$, and reference geocentric time of the first image $t_0$ across all ten systems, as listed in Table~\ref{tab:2}. The geocentric time of the second image is given by $t_1=t_0+\Delta t$, where $\Delta t$ is the median of the EM-derived relative time-delay distribution for each system.

The luminosity distance $d_L$ is determined separately for each system from the spectroscopically measured source redshift $z_s$ listed in Table~\ref{tab:1}, using the adopted cosmological model.

We use the \texttt{IMRPhenomXPHM} waveform model~\citep{IMRPhenomXPHM} to generate the unlensed GW polarizations. Its construction builds on the dominant-mode and higher-mode models of the IMRPhenomX family~\citep{IMRPhenomXAS,IMRPhenomXHM}, available in the LALSuite waveform library\footnote{\url{https://lscsoft.docs.ligo.org/lalsuite/}}. Under the geometric-optics approximation, the total frequency-domain lensed waveform for a system producing $N$ images can be expressed as~\citep{TakahashiNakamura2003,DaiVenumadhav2017,Ezquiaga2021}
\begin{equation}
h^{\rm L}(f)=\sum_{i=0}^{N-1}\sqrt{|\mu_i|}\,e^{-\mathrm{i}\pi(2f\Delta t_i+n_i)}h^{\rm UL}(f),
\label{eq:2}
\end{equation}
where $h^{\rm UL}(f)$ is the unlensed waveform, $\mu_i$ is the signed magnification of the $i$-th image, and $\Delta t_i$ is its time delay relative to the first image ($\Delta t_0=0$). The Morse index $n_i$ takes values of $0$, $1/2$, and $1$ for images corresponding to minima, saddle points, and maxima of the Fermat potential, respectively.

We consider only two-image lensed GW configurations in our simulations and assign the Morse indices $(n_0,n_1)=(0,1/2)$ to the first and second images, respectively. Thus, the corresponding lensed waveforms are given by
\begin{equation}
\begin{aligned}
h_0^{\rm L}(f)&=\sqrt{|\mu_0|}\,h^{\rm UL}(f),\\
h_1^{\rm L}(f)&=\sqrt{|\mu_1|}\,e^{-\mathrm{i}\pi(2f\Delta t+1/2)}h^{\rm UL}(f).
\end{aligned}
\label{eq:3}
\end{equation}

In the actual GW signal injection procedure, the relative time delay is implemented by assigning the geocentric arrival times of the two images as $t_0$ and $t_1=t_0+\Delta t$, respectively, rather than by explicitly applying the frequency-domain time-delay factor to the waveforms. For each lensing system, the two lensed GW images are injected separately into two independent simulated data segments using the three-interferometer Einstein Telescope (ET) configuration implemented in \texttt{Bilby}\footnote{\url{https://bilby-dev.github.io/bilby/}}~\citep{Bilby2019,Bilby2020}. Each data segment has a duration of $8\,\mathrm{s}$ and a sampling frequency of $4096\,\mathrm{Hz}$. Both the waveform reference frequency and the lower-frequency cutoff are set to $10\,\mathrm{Hz}$.

To characterize the signal strength of each simulated lensed GW pair, we calculate the optimal signal-to-noise ratio (SNR) of each image in ET. The network optimal SNR of image $i$ is defined as
\begin{equation}
\rho_i=\sqrt{\sum_{k=1}^{N_{\rm det}}(h_{i,k}|h_{i,k})},
\end{equation}
where $h_{i,k}$ is the GW signal of image $i$ in detector $k$, and $(a|b)$ denotes the noise-weighted inner product,
\begin{equation}
(a|b)=4\,{\rm Re}\int_{f_{\rm min}}^{f_{\rm max}}
\frac{\tilde a(f)\tilde b^*(f)}{S_n(f)}\,df.
\end{equation}
Assuming independent detector noise and separate data segments for the two images, the combined optimal SNR is
\begin{equation}
\rho_c=\sqrt{\rho_0^2+\rho_1^2}.
\label{eq:combined_snr}
\end{equation}
Here, $\rho_0$ and $\rho_1$ are the ET network optimal SNRs of the two lensed images, respectively.

\subsection{Joint parameter estimation}\label{sec:2.3}
In this study, we consider a scenario in which the two GW signals are assumed to be confirmed lensed images of the same BBH merger, and their corresponding EM strong-lensing system has already been identified. Our aim is therefore not to identify lensed GW event pairs or search for their EM counterparts, but to investigate how GW observations can provide additional constraints on the properties of a known strong-lensing system. Under this assumption, we incorporate the available EM information into the GW parameter estimation by fixing $\boldsymbol{\eta}_{\rm E}=(\alpha_{\rm E},\delta_{\rm E},z_{s,\rm E})$, where $(\alpha_{\rm E},\delta_{\rm E})$ are the catalog coordinates of the foreground lens galaxy and $z_{s,\rm E}$ is the spectroscopically determined redshift of the background source galaxy. We approximate the sky directions of both lensed GW images by the foreground lens coordinates, neglecting their arcsecond-scale angular separations. Assuming a fixed cosmological model, the luminosity distance is calculated from the EM-measured source redshift as $d_L=d_L(z_{s,\rm E})$. Fixing the luminosity distance removes the overall magnification--distance degeneracy in the GW amplitude, allowing the absolute magnifications of the two images to be constrained. Throughout the GW analysis, we neglect uncertainties in these EM-provided quantities and treat them as fixed parameters.

EM strong-lens modeling provides posterior constraints on the physical lens parameters, from which the image magnifications and relative time delays can be predicted. However, these derived observables remain subject to uncertainties and correlations in the inferred lens model. Meanwhile, the two lensed GW signals share the same source parameters, while their observed amplitudes and arrival times are modified by gravitational lensing. Joint parameter estimation of the two GW signals therefore provides complementary constraints on the image magnifications and relative time delay. By combining the two types of observations, EM information can further improve the GW parameter constraints, while the GW measurements can be propagated back to the physical lens parameters through the lens mapping, potentially improving the inference of the lens properties.

We first perform joint parameter estimation of the two lensed GW signals and then combine the resulting GW constraints with the EM lens-model posterior. Rather than repeating the EM lens modeling or performing a full joint parameter estimation from scratch, we implement the GW--EM joint inference through posterior reweighting of the existing GW and EM posterior samples.

\subsubsection{Joint parameter estimation of two lensed GW images}
\label{sec:2.3.1}

For each lensing system, we perform a joint parameter estimation of the two lensed GW signals, denoted by $d_0$ and $d_1$. Since the two images originate from the same BBH merger, they share the same source parameters, while their magnifications and geocentric arrival times are treated as image-dependent parameters. We define the sampled parameter set as
\begin{equation}
\boldsymbol{\Theta}=(\mathcal{M},q,\theta_{JN},\phi_c,\mu_0^{\rm GW},\mu_1^{\rm GW},t_0,t_1),
\label{eq:4}
\end{equation}
where $\mathcal{M}$ is the detector-frame chirp mass, $q=m_2/m_1\leq1$ is the mass ratio, $\theta_{JN}$ is the viewing angle, and $\phi_c$ is the coalescence phase. The quantities $\mu_0^{\rm GW}$ and $\mu_1^{\rm GW}$ denote the absolute magnifications of the two images inferred from the GW signals. The GW-inferred relative time delay is derived from the two sampled geocentric arrival times as $\Delta t^{\rm GW}=t_1-t_0$.

For each image, we construct a separate \texttt{GravitationalWaveTransient} likelihood using the corresponding ET data segment and the lensed waveform model described in Sec.~\ref{sec:2.2.2}. Assuming independent stationary Gaussian noise in the two data segments, we use the standard compact-binary likelihood~\citep{CutlerFlanagan1994,Veitch2015}, and the joint likelihood is given by
\begin{equation}
\mathcal{L}_{\rm joint}(d_0,d_1\mid\boldsymbol{\Theta},\boldsymbol{\eta}_{\rm E})=\mathcal{L}_0(d_0\mid\boldsymbol{\Theta},\boldsymbol{\eta}_{\rm E})\mathcal{L}_1(d_1\mid\boldsymbol{\Theta},\boldsymbol{\eta}_{\rm E}),
\label{eq:5}
\end{equation}
or equivalently,
\begin{equation}
\ln\mathcal{L}_{\rm joint}=\ln\mathcal{L}_0+\ln\mathcal{L}_1.
\label{eq:6}
\end{equation}

As described above, the EM-provided parameters $\boldsymbol{\eta}_{\rm E}$ are fixed in the GW analysis. The luminosity distance is calculated from the spectroscopically determined source redshift using the adopted cosmological model and is also held fixed. This removes the overall magnification--distance degeneracy in the GW amplitude, allowing the absolute magnifications of the two images to be constrained. In addition, the spin magnitudes $(a_1,a_2)$, spin orientation angles $(\theta_1,\theta_2,\phi_{12},\phi_{JL})$, and polarization angle $\psi$ are fixed to their respective injection values to reduce the dimensionality of the parameter space.

The priors adopted for the joint parameter estimation are summarized in Table~\ref{tab:3}. We use the \texttt{UniformInComponentsChirpMass} and \texttt{UniformInComponentsMassRatio} priors implemented in \texttt{Bilby} for the detector-frame chirp mass $\mathcal{M}$ and mass ratio $q$, respectively, with additional constraints imposed on the component masses. Independent priors are assigned to the magnifications and geocentric arrival times of the two images, with the arrival-time priors centered on their respective injection values.

According to Bayes' theorem, the joint GW posterior is
\begin{equation}
P(\boldsymbol{\Theta}\mid d_{\rm G},\boldsymbol{\eta}_{\rm E})\propto\mathcal{L}_{\rm joint}(d_{\rm G}\mid\boldsymbol{\Theta},\boldsymbol{\eta}_{\rm E})P_{\rm G}(\boldsymbol{\Theta}\mid\boldsymbol{\eta}_{\rm E}),
\label{eq:7}
\end{equation}
where $d_{\rm G}=\{d_0,d_1\}$ and $P_{\rm G}$ denotes the adopted GW prior. Posterior sampling is performed using the \texttt{dynesty}\footnote{\url{https://dynesty.readthedocs.io/}} nested sampler~\citep{Dynesty2020}, based on nested sampling~\citep{Skilling2006} and implemented in \texttt{Bilby}, with the \texttt{acceptance-walk} sampling method. By marginalizing over the remaining parameters, we obtain the joint GW posterior distribution $P(\mu_0^{\rm GW},\mu_1^{\rm GW},\Delta t^{\rm GW}\mid d_{\rm G},\boldsymbol{\eta}_{\rm E})$, which is subsequently used in the GW--EM posterior reweighting.

\subsubsection{GW--EM joint inference through posterior reweighting}
\label{sec:2.3.2}

Having obtained the joint GW constraints on the two image magnifications and relative time delay in Sec.~\ref{sec:2.3.1}, we now combine them with the EM lens-model posterior within a unified Bayesian framework. We denote the GW and EM datasets by $d_{\rm G}=\{d_0,d_1\}$ and $d_{\rm E}$, respectively. Let $\boldsymbol{\theta}$ represent the GW parameters other than the lensing observables, including the reference geocentric arrival time, and let $\boldsymbol{\lambda}$ denote the lens-model parameters defined in Eq.~(\ref{eq:1}). The two observations are connected through the lens mapping $\boldsymbol{\gamma}=g(\boldsymbol{\lambda})=(|\mu_0|,|\mu_1|,\Delta t)$, as defined in Sec.~\ref{sec:2.2.1}.

Assuming that the GW and EM observations are conditionally independent given the model parameters, we construct a joint prior from the EM lens-parameter prior and the conditional GW prior for the remaining GW parameters. The latter is written as $P_{\rm G}(\boldsymbol{\theta}\mid\boldsymbol{\gamma},\boldsymbol{\eta}_{\rm E})$ to retain any prior dependence between the GW nuisance parameters and the lensing observables. The joint posterior conditioned on the fixed EM information $\boldsymbol{\eta}_{\rm E}$ can then be expressed as
\begin{equation}
\begin{aligned}
&P(\boldsymbol{\theta},\boldsymbol{\lambda}\mid d_{\rm G},d_{\rm E},\boldsymbol{\eta}_{\rm E})\\
&\quad=\frac{1}{\mathcal{Z}_{\rm GE}}
P(d_{\rm G}\mid\boldsymbol{\theta},g(\boldsymbol{\lambda}),\boldsymbol{\eta}_{\rm E})\\
&\qquad\times P(d_{\rm E}\mid\boldsymbol{\lambda},\boldsymbol{\eta}_{\rm E})
P_{\rm G}(\boldsymbol{\theta}\mid g(\boldsymbol{\lambda}),\boldsymbol{\eta}_{\rm E})
P_{\rm E}(\boldsymbol{\lambda}\mid\boldsymbol{\eta}_{\rm E}),
\end{aligned}
\label{eq:8}
\end{equation}
where $P_{\rm G}$ and $P_{\rm E}$ denote the adopted GW and EM priors, respectively, and $\mathcal{Z}_{\rm GE}$ is the joint evidence.

The EM-only lens-parameter posterior satisfies
\begin{equation}
P(\boldsymbol{\lambda}\mid d_{\rm E},\boldsymbol{\eta}_{\rm E})\propto P(d_{\rm E}\mid\boldsymbol{\lambda},\boldsymbol{\eta}_{\rm E})P_{\rm E}(\boldsymbol{\lambda}\mid\boldsymbol{\eta}_{\rm E}).
\label{eq:9}
\end{equation}
Substituting this relation into Eq.~(\ref{eq:8}), we obtain
\begin{equation}
\begin{aligned}
&P(\boldsymbol{\theta},\boldsymbol{\lambda}\mid d_{\rm G},d_{\rm E},\boldsymbol{\eta}_{\rm E})\\
&\quad\propto P(d_{\rm G}\mid\boldsymbol{\theta},g(\boldsymbol{\lambda}),\boldsymbol{\eta}_{\rm E})\\
&\qquad\times P_{\rm G}(\boldsymbol{\theta}\mid g(\boldsymbol{\lambda}),\boldsymbol{\eta}_{\rm E})
P(\boldsymbol{\lambda}\mid d_{\rm E},\boldsymbol{\eta}_{\rm E}).
\end{aligned}
\label{eq:10}
\end{equation}

This expression provides the basis for combining the GW and EM constraints without repeating the EM lens modeling. In practice, we implement the joint inference through posterior reweighting, using the existing GW and EM posterior samples to incorporate the additional likelihood information provided by the complementary dataset.

We first incorporate the EM constraints into the existing GW posterior. The EM lens-model posterior induces a posterior distribution of the lensing observables through the mapping $\boldsymbol{\gamma}=g(\boldsymbol{\lambda})$, which can be expressed as
\begin{equation}
P(\boldsymbol{\gamma}\mid d_{\rm E},\boldsymbol{\eta}_{\rm E})=\int \delta_{\rm D}^{(3)}[\boldsymbol{\gamma}-g(\boldsymbol{\lambda})]P(\boldsymbol{\lambda}\mid d_{\rm E},\boldsymbol{\eta}_{\rm E})\,d\boldsymbol{\lambda}.
\label{eq:11}
\end{equation}
Here, $\delta_{\rm D}^{(3)}$ denotes the three-dimensional Dirac delta function. Similarly, the corresponding EM-induced prior is given by
\begin{equation}
P_{\rm E}(\boldsymbol{\gamma}\mid\boldsymbol{\eta}_{\rm E})=\int \delta_{\rm D}^{(3)}[\boldsymbol{\gamma}-g(\boldsymbol{\lambda})]P_{\rm E}(\boldsymbol{\lambda}\mid\boldsymbol{\eta}_{\rm E})\,d\boldsymbol{\lambda}.
\label{eq:12}
\end{equation}
According to Bayes' theorem, the ratio of these two distributions is proportional to the additional EM likelihood information in the common observable space:
\begin{equation}
\mathcal{R}_{\rm E}(\boldsymbol{\gamma})=\frac{P(\boldsymbol{\gamma}\mid d_{\rm E},\boldsymbol{\eta}_{\rm E})}{P_{\rm E}(\boldsymbol{\gamma}\mid\boldsymbol{\eta}_{\rm E})}.
\label{eq:13}
\end{equation}
The density ratio is evaluated within the support of the corresponding EM-induced prior.

The existing GW posterior can therefore be reweighted using the additional EM constraints. For each GW posterior sample $\boldsymbol{\Theta}^{(k)}$, we extract the corresponding lensing observables $\boldsymbol{\gamma}^{(k)}=(\mu_0^{{\rm GW},(k)},\mu_1^{{\rm GW},(k)},\Delta t^{{\rm GW},(k)})$ and assign an importance weight
\begin{equation}
w_{{\rm G},k}\propto\mathcal{R}_{\rm E}(\boldsymbol{\gamma}^{(k)})=\frac{P(\boldsymbol{\gamma}^{(k)}\mid d_{\rm E},\boldsymbol{\eta}_{\rm E})}{P_{\rm E}(\boldsymbol{\gamma}^{(k)}\mid\boldsymbol{\eta}_{\rm E})}.
\label{eq:14}
\end{equation}
Here, the subscript G indicates that the weights are assigned to the original GW posterior samples. The resulting weighted GW posterior incorporates the additional information provided by the EM observations while retaining the original GW posterior samples.

Conversely, we use the GW constraints to update the EM lens-parameter posterior. Marginalizing Eq.~(\ref{eq:10}) over the GW parameters $\boldsymbol{\theta}$ yields
\begin{equation}
P(\boldsymbol{\lambda}\mid d_{\rm G},d_{\rm E},\boldsymbol{\eta}_{\rm E})\propto P(\boldsymbol{\lambda}\mid d_{\rm E},\boldsymbol{\eta}_{\rm E})\mathcal{L}_{\rm G}^{\rm marg}(g(\boldsymbol{\lambda})),
\label{eq:15}
\end{equation}
where the marginalized GW likelihood is defined as
\begin{equation}
\mathcal{L}_{\rm G}^{\rm marg}(\boldsymbol{\gamma})=\int P(d_{\rm G}\mid\boldsymbol{\theta},\boldsymbol{\gamma},\boldsymbol{\eta}_{\rm E})P_{\rm G}(\boldsymbol{\theta}\mid\boldsymbol{\gamma},\boldsymbol{\eta}_{\rm E})\,d\boldsymbol{\theta}.
\label{eq:16}
\end{equation}
The conditional prior is retained in this expression because the original GW sampling priors may induce correlations between the reference arrival time and the relative time delay. Using Bayes' theorem, the marginalized GW likelihood can be recovered, up to a normalization constant, from the GW posterior and its corresponding induced prior:
\begin{equation}
\mathcal{L}_{\rm G}^{\rm marg}(\boldsymbol{\gamma})\propto\frac{P(\boldsymbol{\gamma}\mid d_{\rm G},\boldsymbol{\eta}_{\rm E})}{P_{\rm G}(\boldsymbol{\gamma}\mid\boldsymbol{\eta}_{\rm E})},
\label{eq:17}
\end{equation}
where $P_{\rm G}(\boldsymbol{\gamma}\mid\boldsymbol{\eta}_{\rm E})$ is the marginal prior on the lensing observables induced by the original GW sampling priors.

Substituting Eq.~(\ref{eq:17}) into Eq.~(\ref{eq:15}), we obtain
\begin{equation}
\begin{aligned}
&P(\boldsymbol{\lambda}\mid d_{\rm G},d_{\rm E},\boldsymbol{\eta}_{\rm E})\\
&\quad\propto P(\boldsymbol{\lambda}\mid d_{\rm E},\boldsymbol{\eta}_{\rm E})
\frac{P(g(\boldsymbol{\lambda})\mid d_{\rm G},\boldsymbol{\eta}_{\rm E})}{P_{\rm G}(g(\boldsymbol{\lambda})\mid\boldsymbol{\eta}_{\rm E})}.
\end{aligned}
\label{eq:18}
\end{equation}
Accordingly, for each EM posterior sample $\boldsymbol{\lambda}^{(k)}$, we evaluate the corresponding lensing observables $\boldsymbol{\gamma}^{(k)}=g(\boldsymbol{\lambda}^{(k)})$ and assign an importance weight
\begin{equation}
w_{{\rm E},k}\propto\frac{P(\boldsymbol{\gamma}^{(k)}\mid d_{\rm G},\boldsymbol{\eta}_{\rm E})}{P_{\rm G}(\boldsymbol{\gamma}^{(k)}\mid\boldsymbol{\eta}_{\rm E})}.
\label{eq:19}
\end{equation}
Here, the subscript E indicates that the weights are assigned to the original EM posterior samples. The reweighted EM posterior allows the GW constraints on the image magnifications and relative time delay to be propagated back to the physical lens parameters.

The two reweighting procedures incorporate the additional likelihood information from the complementary dataset into their respective original posterior distributions. Their resulting joint posterior distributions are strictly equivalent only when the underlying prior specifications are mutually consistent. In the present analysis, the two directions are therefore used to quantify the improvements relative to their respective original GW and EM posteriors.

In practice, the probability densities required for the importance weights are estimated using Gaussian kernel density estimation (KDE)~\citep{Silverman1986KDE}. To account for the broad dynamic range of the image magnifications, we perform the density estimation in the transformed parameter space
\begin{equation}
\boldsymbol{x}=(\ln|\mu_0|,\ln|\mu_1|,\Delta t).
\label{eq:20}
\end{equation}
For the GW posterior samples, the corresponding variables are $\boldsymbol{x}^{\rm GW}=(\ln\mu_0^{\rm GW},\ln\mu_1^{\rm GW},\Delta t^{\rm GW})$. The EM posterior and prior samples are mapped into the same parameter space, allowing the density ratios in Eqs.~(\ref{eq:14}) and \eqref{eq:19} to be evaluated consistently. Since each posterior density and its corresponding prior density are evaluated in the same transformed space, the Jacobian factors cancel in their ratio.

Special care is taken when constructing the GW prior distribution of the relative time delay. In the GW parameter estimation, the two geocentric arrival times are assigned independent uniform priors centered on their respective injection values, each with a half-width of $a=0.1\,\mathrm{s}$. Consequently, the induced marginal prior on $\Delta t^{\rm GW}=t_1-t_0$ follows a triangular distribution,
\begin{equation}
\begin{aligned}
&P_{\rm G}(\Delta t^{\rm GW}\mid\boldsymbol{\eta}_{\rm E})\\
&=
\begin{cases}
\dfrac{2a-|\Delta t^{\rm GW}-\Delta t_{\rm inj}|}{4a^2},\qquad |\Delta t^{\rm GW}-\Delta t_{\rm inj}|\leq 2a,\\[6pt]
0,\quad \text{otherwise}.
\end{cases}
\end{aligned}
\label{eq:21}
\end{equation}
where $\Delta t_{\rm inj}=t_{1,\rm inj}-t_{0,\rm inj}$. This induced prior is included when evaluating the GW posterior-to-prior density ratio, together with the corresponding magnification priors.

For each reweighting direction, the resulting importance weights are normalized according to
\begin{equation}
\widetilde{w}_k=\frac{w_k}{\sum_{j=1}^{N}w_j},
\label{eq:22}
\end{equation}
where $N$ denotes the number of original posterior samples. The statistical efficiency of the reweighting is quantified using the effective sample size (ESS)~\citep{Vehtari2024PSIS,Janquart2023GOLUM},
\begin{equation}
N_{\rm eff}=\frac{1}{\sum_{k=1}^{N}\widetilde{w}_k^2}.
\label{eq:23}
\end{equation}
The reweighted posterior distributions are then used to calculate the median values and $90\%$ credible intervals of the parameters of interest. In particular, the EM-reweighted GW samples provide updated constraints on the GW parameters and lensing observables, while the GW-reweighted EM samples provide updated constraints on the physical lens parameters.

The relative time delay can be measured with high precision from the two GW signals, whereas the time-delay distribution predicted by the EM lens model is generally much broader. Directly applying such a narrow GW constraint to a finite set of EM posterior samples may cause the importance weights to concentrate on only a few samples, resulting in a substantially reduced effective sample size. Furthermore, the EM-predicted time delay may be affected by systematic uncertainties associated with the adopted lens model that are not fully captured by its statistical posterior. A similar issue was addressed by \citet{Uronen2026JointBayesian}, who introduced an additional Gaussian time-delay modeling uncertainty to account for the limited accuracy of EM lens models relative to the precision of GW timing measurements. To examine the sensitivity of the GW-to-EM reweighting to these effects, we additionally introduce an effective time-delay uncertainty defined by
\begin{equation}
\sigma_{t,\rm eff}=\sqrt{\sigma_{t,\rm GW}^2+\sigma_{t,\rm model}^2},
\label{eq:24}
\end{equation}
where $\sigma_{t,\rm GW}$ characterizes the uncertainty inferred from the prior-corrected GW time-delay posterior, and $\sigma_{t,\rm model}$ represents an additional time-delay modeling uncertainty. The two uncertainty contributions are assumed to be independent and are combined in quadrature under a Gaussian approximation. Following the motivation of \citet{Uronen2026JointBayesian}, we account for additional time-delay modeling uncertainty in the reweighting procedure. Unlike their formulation, which introduces a Gaussian modeling uncertainty in the effective GW time-delay likelihood, we explicitly combine the uncertainty inferred from the prior-corrected GW time-delay posterior with the additional modeling uncertainty in quadrature.

In this alternative implementation, we incorporate the magnification constraints and the broadened time-delay constraint separately. Specifically, the importance weight assigned to each EM posterior sample is approximated as
\begin{equation}
\begin{aligned}
w_{{\rm E},k}^{\rm alt}
&\propto\mathcal{R}_{\rm G}^{(\mu)}(\boldsymbol{\mu}^{(k)})\\
&\quad\times\exp\left[
-\frac{(\Delta t^{(k)}-\widehat{\Delta t}_{\rm GW})^2}
{2\sigma_{t,\rm eff}^2}
\right],
\end{aligned}
\label{eq:25}
\end{equation}
where $\boldsymbol{\mu}^{(k)}=(|\mu_0^{(k)}|,|\mu_1^{(k)}|)$, and $\mathcal{R}_{\rm G}^{(\mu)}$ denotes the GW posterior-to-prior density ratio evaluated in the two-dimensional logarithmic magnification space.

The central time delay $\widehat{\Delta t}_{\rm GW}$ and its uncertainty $\sigma_{t,\rm GW}$ are estimated from the prior-corrected GW time-delay posterior, using weights proportional to $1/P_{\rm G}(\Delta t^{\rm GW}\mid\boldsymbol{\eta}_{\rm E})$. The resulting weighted mean and standard deviation are used to construct a Gaussian approximation to the marginalized GW time-delay likelihood. An additional modeling uncertainty $\sigma_{t,\rm model}$ is then incorporated through Eq.~(\ref{eq:24}). This alternative formulation approximates the magnification and time-delay constraints as separate contributions and is therefore not strictly equivalent to the full three-dimensional KDE reweighting. By varying $\sigma_{t,\rm model}$, we assess the sensitivity of the reweighted EM posterior to the assumed time-delay modeling uncertainty.

\begin{table}[tbp]
\centering
\caption{Priors adopted for the joint parameter estimation of the two lensed GW images. The spin parameters and polarization angle are fixed to their injection values given in Table~\ref{tab:2}, while the sky location and source redshift are fixed using the corresponding EM information.}
\label{tab:3}
\small
\begin{tabular*}{\columnwidth}
{@{\extracolsep{\fill}}lll@{}}
\toprule
Parameter & Prior & Range \\
\midrule
$\mathcal M$
& Uniform in component masses
& $[25,200]\,\Msun$ \\
$q$
& Uniform in component masses
& $[0.125,1]$ \\
$m_1,m_2$
& Constrained
& $[5,200]\,\Msun$ \\
$\theta_{JN}$
& Sine
& $[0,\pi]$ \\
$\phi_c$
& Uniform
& $[0,2\pi]$ \\
$t_i$
& Uniform
& $t_{i,\rm inj}\pm0.1\,\mathrm{s}$ \\
$\mu_i$
& Log-uniform
& $[0.1,50]$ \\
\bottomrule
\end{tabular*}
\end{table}

\section{Results}\label{sec:3}
\subsection{Joint parameter estimation of the two lensed GW images}
\label{sec:3.1}

\begin{figure*}[t]
\centering
\includegraphics[width=\textwidth]{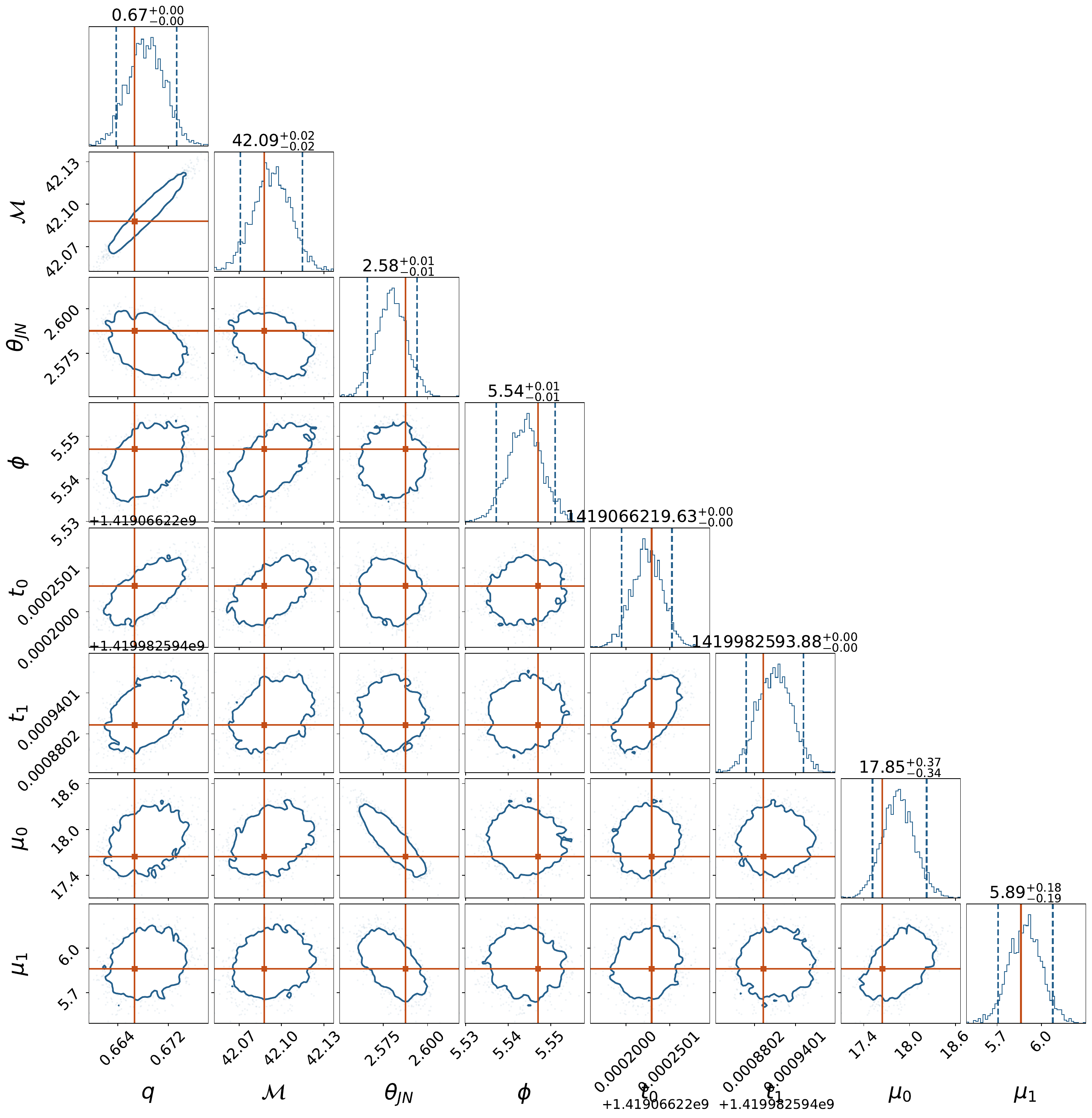}
\caption{Joint posterior distributions obtained from the two-image GW parameter estimation for the J0252+0039 lensing system. The sampled parameters are the mass ratio $q$, detector-frame chirp mass $\mathcal{M}$, viewing angle $\theta_{JN}$, coalescence phase $\phi_c$, geocentric arrival times $(t_0,t_1)$, and absolute magnifications $(\mu_0^{\rm GW},\mu_1^{\rm GW})$. The diagonal panels show the marginalized one-dimensional posterior distributions, while the off-diagonal panels show the corresponding two-dimensional posterior contours enclosing $90\%$ of the posterior probability. The orange lines and markers indicate the injection values, and the dashed blue lines mark the central $90\%$ credible intervals.}
\label{fig.PE_0039}
\end{figure*}

We first examine the parameter constraints obtained from the joint analysis of the two lensed GW signals before incorporating the EM lens-model posterior through reweighting. Fig.~\ref{fig.PE_0039} presents the joint posterior distributions for the J0252+0039 system as a representative example. The eight sampled parameters include the detector-frame chirp mass $\mathcal{M}$, mass ratio $q$, viewing angle $\theta_{JN}$, coalescence phase $\phi_c$, geocentric arrival times $(t_0,t_1)$, and absolute magnifications $(\mu_0^{\rm GW},\mu_1^{\rm GW})$.

An important feature of the present analysis is that the source redshift is provided by the identified EM strong-lensing system. Assuming the adopted cosmological model, the corresponding luminosity distance is calculated from the EM-measured source redshift and held fixed throughout the GW parameter estimation. Since the observed GW amplitude of each lensed image scales as $\sqrt{|\mu_i|}/d_L$, fixing $d_L$ removes the overall magnification--distance degeneracy and allows the absolute magnifications of both images to be inferred directly from the GW signals. This differs from a GW-only analysis with an unknown source distance, in which the amplitudes primarily constrain the effective luminosity distances and the relative magnification between the images. Nevertheless, correlations between the absolute magnifications and other amplitude-related source parameters, particularly the viewing angle, remain.

As shown in Fig.~\ref{fig.PE_0039}, the injected values are consistent with the recovered posterior distributions, demonstrating that the joint analysis successfully recovers the simulated two-image GW signals under the adopted waveform model and parameter assumptions. The posterior distributions show that the binary mass parameters, represented by $\mathcal{M}$ and $q$, are relatively well constrained. The two geocentric arrival times are also tightly constrained, allowing the relative time delay $\Delta t^{\rm GW}=t_1-t_0$ to be determined with high precision.

In contrast, the absolute magnification posteriors exhibit broader uncertainties and noticeable correlations with the viewing angle $\theta_{JN}$. In particular, the tilted contours in the $(\theta_{JN},\mu_0^{\rm GW})$ and $(\theta_{JN},\mu_1^{\rm GW})$ planes reflect the partial degeneracy between the source orientation and the lensing-induced amplitude enhancement. Consequently, although fixing the luminosity distance enables the inference of the absolute magnifications, their measurement precision remains affected by the uncertainty in the source orientation.

These results provide the baseline for evaluating the additional information supplied by the EM lens model. In the following subsection, we investigate how incorporating the EM-derived constraints on the image magnifications and relative time delay modifies the GW posterior distributions, particularly those of the absolute magnifications and viewing angle.

\subsection{Improved GW constraints through EM reweighting}
\label{sec:3.2}

\begin{figure}[tbp]
\centering
\includegraphics[width=\linewidth]{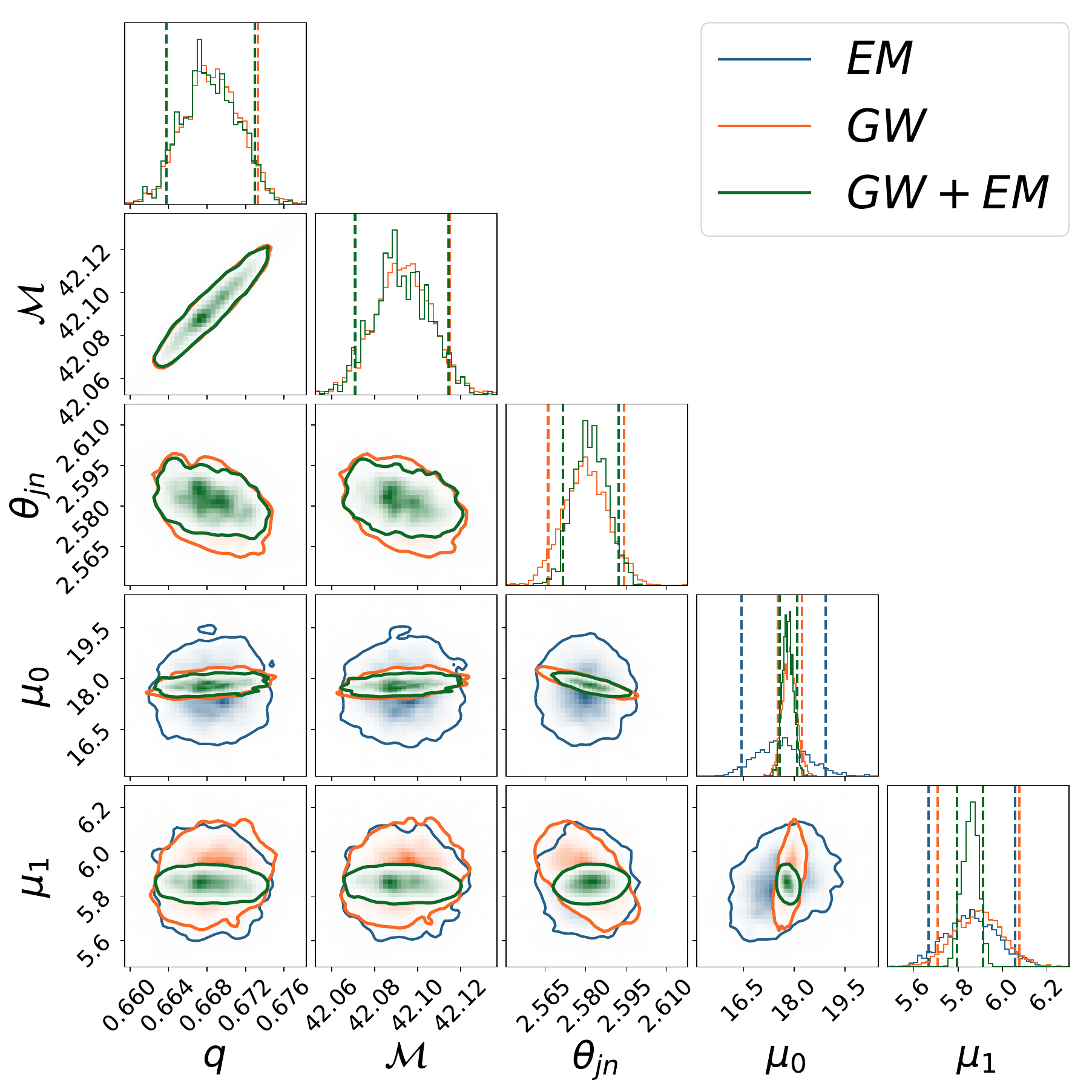}
\caption{Illustrative posterior comparison for J0252+0039, showing the mass ratio $q$, detector-frame chirp mass $\mathcal M$ (in solar masses), viewing angle $\theta_{JN}$ (in radians), and absolute magnifications $(\mu_0,\mu_1)$. Orange denotes the GW posterior conditioned on the fixed external information $\eta_{\rm EM}$, and green includes the additional EM lens-model information. Blue represents EM-only constraints in the magnification panels. The blue mass and viewing-angle distributions are display substitutes drawn from the GW posterior, not EM constraints; they are independently paired with EM magnification samples. Blue cross-panels involving these substitutes have no inferential meaning and do not represent EM parameter correlations. The orange and green contours show the GW correlations through which EM magnification information affects the viewing-angle constraint. The two-dimensional contours enclose $90\%$ of the corresponding joint posterior probability. Vertical dashed lines mark the $90\%$ credible-interval limits.}
\label{fig:gwcorner}
\end{figure}

\begin{figure}[tbp]
\centering
\includegraphics[width=\linewidth]{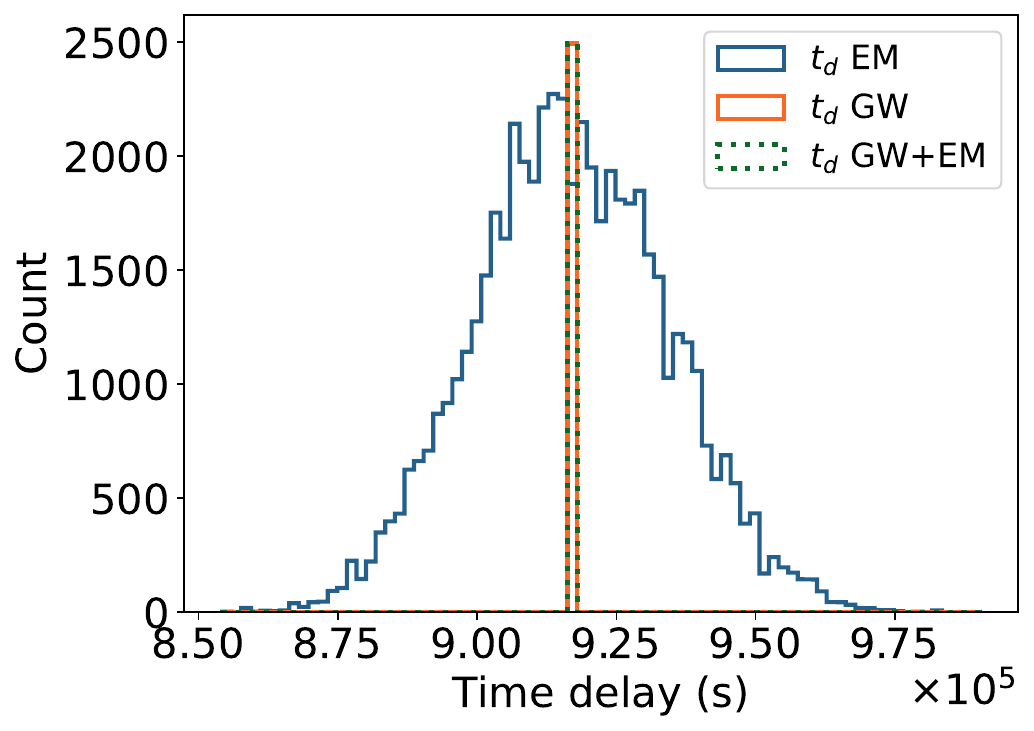}
\caption{Illustrative distributions of the relative delay $\Delta t$ inferred from the EM lens model (blue), the conditional two-image GW analysis (orange), and the EM-informed GW update (green). The horizontal axis is in seconds. The GW and updated distributions nearly overlap and are much narrower than the EM prediction, indicating that the time-delay constraint is dominated by GW data.}
\label{fig:delay}
\end{figure}

We next investigate how the EM lens-model information improves the parameter constraints obtained from the joint two-image GW analysis. Using the posterior reweighting procedure described in Sec.~\ref{sec:2.3.2}, we incorporate the EM-derived constraints on the absolute magnifications and relative time delay into the GW posterior. Fig.~\ref{fig:gwcorner} compares the conditional GW posterior (orange) with the EM-reweighted GW posterior (green) for J0252+0039. The blue magnification distributions are obtained from the EM lens model. For display only, the blue distributions of $q$, $\mathcal M$, and $\theta_{JN}$ are drawn from GW posterior samples and independently paired with EM magnification samples. They are not EM measurements, and the associated blue cross-panels do not encode physical EM constraints or parameter correlations. The interpretation of information transfer between source parameters and magnifications is based on the orange and green posteriors.

The most noticeable improvement occurs in the absolute magnifications. As shown in Fig.~\ref{fig:gwcorner}, the EM-only magnification distributions provide complementary information to the GW measurements, resulting in narrower posterior distributions for both $\mu_0^{\rm GW}$ and $\mu_1^{\rm GW}$. The improvement is also visible in the corresponding two-dimensional posterior contours. In contrast, the EM observations do not directly constrain the intrinsic BBH mass parameters or the viewing angle. Nevertheless, incorporating the EM magnification information also narrows the posterior of $\theta_{JN}$. This improvement arises from the correlations between the viewing angle and the absolute magnifications in the GW posterior, through which the additional EM information is propagated to the source-orientation constraint.

To quantify these improvements, we define the percentage reduction in the central $90\%$ credible interval width of a parameter $x$ as
\begin{equation}
\mathcal{R}_x^{(G)}=
\left(1-\frac{W_x^{\rm GW+EM}}{W_x^{\rm GW}}\right)\times100\%,
\label{eq:reduction}
\end{equation}
where $W_x=x_{95}-x_5$ denotes the difference between the 95th and 5th percentiles of the corresponding posterior distribution. For the J0252+0039 system, the credible interval widths of $\mu_0^{\rm GW}$ and $\mu_1^{\rm GW}$ decrease by $26.3\%$ and $68.0\%$, respectively, while that of $\theta_{JN}$ decreases by $26.8\%$. The mass parameters are less affected, with reductions of $3.5\%$ and $2.0\%$ for $q$ and $\mathcal{M}$, respectively. These results demonstrate that the additional EM information primarily improves the magnification constraints and can further enhance the inference of correlated GW source parameters, particularly the viewing angle.

The effect of EM reweighting on the relative time delay is comparatively small. Fig.~\ref{fig:delay} compares the EM-derived, GW-only, and EM-reweighted GW distributions of $\Delta t$. The GW-only and updated distributions nearly overlap and are substantially narrower than the EM prediction, indicating that the relative time delay is already tightly constrained by the two GW signals. Consequently, the additional EM information produces a negligible reduction in its credible interval width. Conversely, when the GW constraints are compared with the original EM predictions, the credible interval widths of $\mu_0$, $\mu_1$, and $\Delta t$ are reduced by $78.8\%$, $69.9\%$, and approximately $100.0\%$, respectively. The last value reflects rounding of an extremely narrow but finite time-delay interval.

Overall, these results highlight the complementary roles of the two observations. The EM lens model provides additional information that improves the GW magnification constraints and, through their correlations, the viewing-angle constraint. Meanwhile, the GW signals provide a substantially more precise measurement of the relative time delay. The improvements across the ten lensing systems are further summarized in Table~\ref{tab:ensemble}.

\begin{table*}[tbp]
\centering
\caption{Reduction in the central 90\% credible-interval widths for the ten strong-lensing systems. The combined optimal SNR $\rho_c$ is defined in Eq.~(\ref{eq:combined_snr}). The $\mathcal R^{(G)}$ columns compare the GW-only and EM-reweighted GW posteriors, while the $\mathcal R^{(E)}$ columns compare the EM-only and GW-reweighted EM posteriors using magnification-only reweighting. Here, $y$ denotes the normalized source--lens offset, and $\Delta r$ denotes the total angular offset between the mass and light centers of the lens galaxy. All reductions are expressed as percentages.}
\label{tab:ensemble}
\small
\setlength{\tabcolsep}{3pt}
\begin{tabular*}{\textwidth}{@{\extracolsep{\fill}}lccccccccc@{}}
\toprule
System
& $\rho_c$
& $\mathcal R_{\mu_0}^{(G)}$
& $\mathcal R_{\mu_1}^{(G)}$
& $\mathcal R_{\theta_{JN}}^{(G)}$
& $\mathcal R_{\theta_E}^{(E)}$
& $\mathcal R_y^{(E)}$
& $\mathcal R_{\Delta r}^{(E)}$
& $\mathcal R_e^{(E)}$
& $\mathcal R_{\gamma_{\rm ext}}^{(E)}$
\\
\midrule
J0252+0039    & 369 & 26.3 & 68.0 & 26.8 &  5.4 & 36.0 & 20.5 & 12.0 & 12.6\\
J0822+2652    & 217 &  9.4 & 35.4 & 10.9 & 10.8 & 56.8 & 54.0 & 29.7 & 14.5\\
J0903+4116    & 147 & 73.4 & 49.8 & 44.4 & -0.5 & 19.7 &  0.1 &  2.1 &  0.2\\
J1023+4230    & 377 & 37.3 & 36.4 & 29.6 &  6.8 & 35.1 &  2.3 & 28.8 & 10.5\\
J1106+5228    & 610 &  5.0 & 58.8 &  7.3 &  0.6 & 61.0 & 33.6 & 35.5 &  4.2\\
J1153+4612    & 106 & 95.3 & 21.0 & 52.2 & -1.1 & 23.0 & 16.2 & 15.7 & 15.1\\
J1204+0358    & 301 & 45.8 & 30.8 & 30.1 &  8.8 & 40.9 & 11.4 & 14.1 &  1.2\\
J1627$-$0053  & 615 & 15.1 & 25.0 & 11.8 &  2.4 & 40.0 & 19.7 &  3.6 & -1.5\\
J2238$-$0754  & 286 & 82.4 & 28.1 & 23.0 &  2.8 & 20.1 & 22.6 &  7.0 &  3.2\\
J2300+0022    & 424 & 50.0 &  7.9 & 11.8 &  8.0 & 62.3 & 51.9 & 40.5 & 21.5\\
\bottomrule
\end{tabular*}
\end{table*}

\subsection{Improved lens-parameter constraints through GW reweighting}
\label{sec:3.3}

We now investigate how the GW measurements can improve the inference of the physical lens parameters through posterior reweighting. For the J0252+0039 system, we first consider the constraints provided by the two absolute GW magnifications alone, before examining the additional effect of the relative time-delay measurement. This allows us to distinguish the information supplied by the GW amplitudes from that introduced by the much more precise GW timing measurement.

To examine how the GW magnification constraints affect the geometrical properties of the lensing system, we consider both the relative source--lens geometry and the mass--light centroid offset of the lens galaxy. For the former, we define the normalized source offsets as
\begin{equation}
y_x=\frac{\beta_x-x_{\rm L}}{\theta_{\rm E}},\qquad
y_y=\frac{\beta_y-y_{\rm L}}{\theta_{\rm E}},
\end{equation}
and their radial combination as
\begin{equation}
y=\sqrt{y_x^2+y_y^2}.
\end{equation}
For the latter, we define the mass--light centroid offsets as
\begin{equation}
\Delta x=x_{\rm L}-x_c,\qquad
\Delta y=y_{\rm L}-y_c,
\end{equation}
where $(x_c,y_c)$ denotes the light center of the lens galaxy. The total angular offset is given by
\begin{equation}
\Delta r=\sqrt{\Delta x^2+\Delta y^2}.
\end{equation}
While $(y_x,y_y,y)$ are dimensionless, $(\Delta x,\Delta y,\Delta r)$ are expressed in arcseconds. In addition, we consider the ellipticity amplitude $e=\sqrt{e_1^2+e_2^2}$ and external-shear amplitude $\gamma_{\rm ext}=\sqrt{\gamma_1^2+\gamma_2^2}$.

To quantify the improvement in lens-parameter constraints, we use Eq.~(\ref{eq:reduction}) with the EM-only credible-interval width as the reference, giving $\mathcal R_x^{(E)}=[1-W_x^{\rm EM+GW}/W_x^{\rm EM}]\times100\%$.

\begin{figure}[tbp]
\centering
\includegraphics[width=\linewidth]{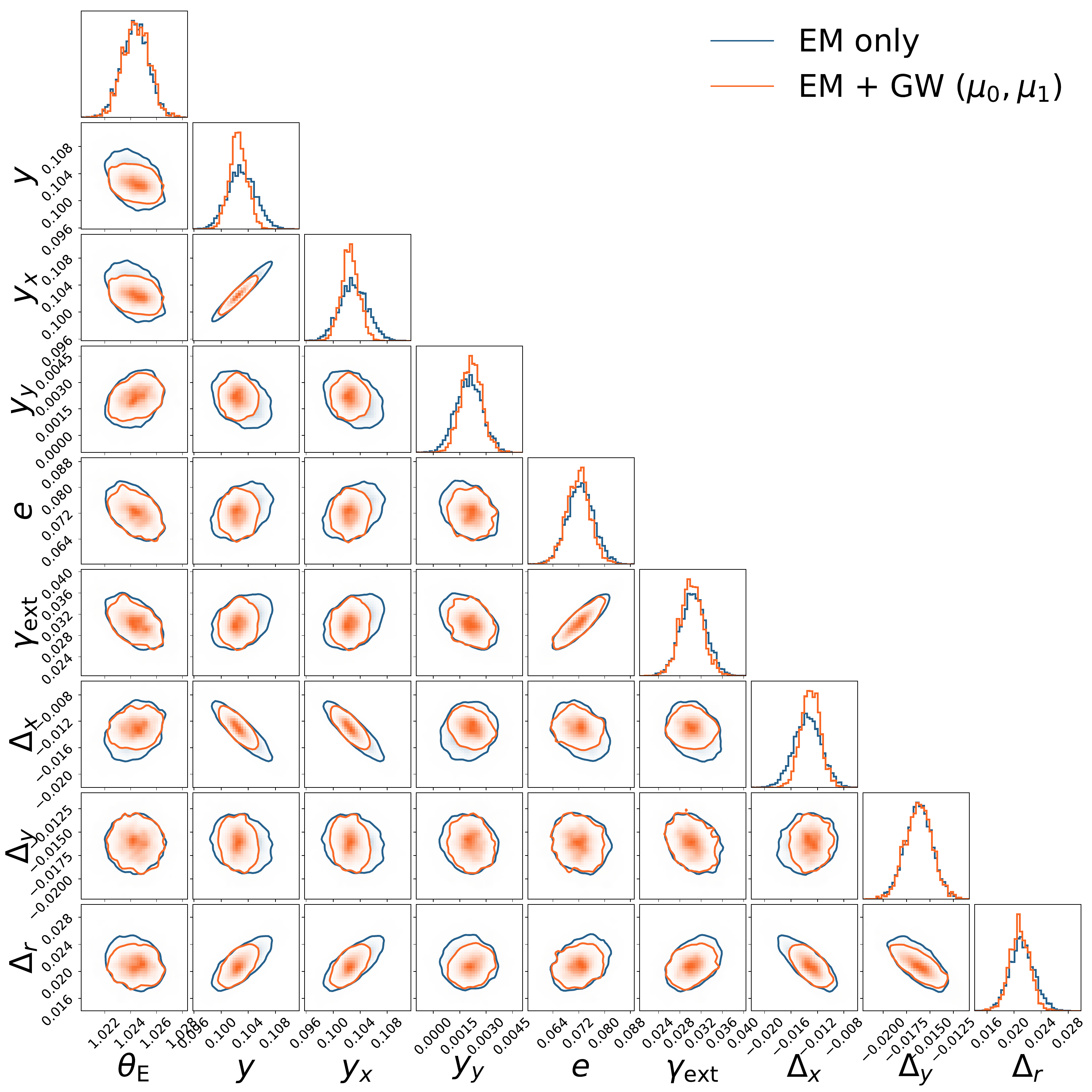}
\caption{
EM-only (blue) and EM+GW (orange) posterior distributions for system J0252+0039, obtained by reweighting the EM posterior samples using the two GW magnification constraints, without time-delay information. The displayed quantities are the Einstein radius $\theta_{\rm E}$, normalized source--lens offsets $(y,y_x,y_y)$, ellipticity amplitude $e$, external-shear amplitude $\gamma_{\rm ext}$, and mass--light centroid offsets $(\Delta x,\Delta y)$, which describe the relative displacement between the lens mass center and light center along the two coordinate axes. All quantities are dimensionless except $\theta_{\rm E}$, $\Delta x$, and $\Delta y$, which are in arcseconds. The clearest narrowing occurs in $y$ and $y_x$, indicating improved constraints on the relative source--lens geometry within the adopted lens model. Their improvements are closely related because $y\simeq|y_x|$ for this system. The two-dimensional contours enclose $90\%$ of the corresponding joint posterior probability.
}
\label{fig:geometry}
\end{figure}

Fig.~\ref{fig:geometry} compares the EM-only posterior with the posterior reweighted using the two GW magnifications in terms of these geometric and amplitude parameters. The clearest narrowing occurs in $y$ and $y_x$, whereas the Einstein radius, ellipticity amplitude, and external-shear amplitude exhibit smaller changes. For J0252+0039, the central $90\%$ credible-interval width of $y$ decreases by $36.0\%$, compared with $5.4\%$, $12.0\%$, and $12.6\%$ for $\theta_{\rm E}$, $e$, and $\gamma_{\rm ext}$, respectively (Table~\ref{tab:ensemble}). The credible-interval width of the mass--light centroid offset $\Delta r$ also decreases by $20.5\%$, indicating that the GW magnification information improves the constraints not only on the relative source--lens geometry but also on the relative positions of the mass and light centers of the lens galaxy. The improvements can be understood from the sensitivity of GW observations to the signal amplitude. Once the source redshift is known and a cosmological model is assumed, the luminosity distance is determined, allowing the absolute magnifications to be constrained from the observed GW amplitudes. For sufficiently high-SNR signals, the GW magnification measurements can provide informative constraints that preferentially select lens-model configurations consistent with the observed magnifications. This additional information is particularly effective for geometric parameters that are strongly correlated with the image magnifications, such as the normalized source--lens offset $y$ and the mass--light centroid offset $\Delta r$. Consequently, incorporating GW magnification information can improve the constraints on both the relative source--lens geometry and the spatial alignment between the mass and light distributions of the lens galaxy. The GW magnification measurements therefore provide complementary information about the relative position of the background source and the foreground lens. The changes in the constraints on the original lens-model parameters are quantified separately by the green bars in Fig.~\ref{fig:three_reduction}. 

\begin{figure*}[tbp]
\centering
\includegraphics[width=\textwidth]{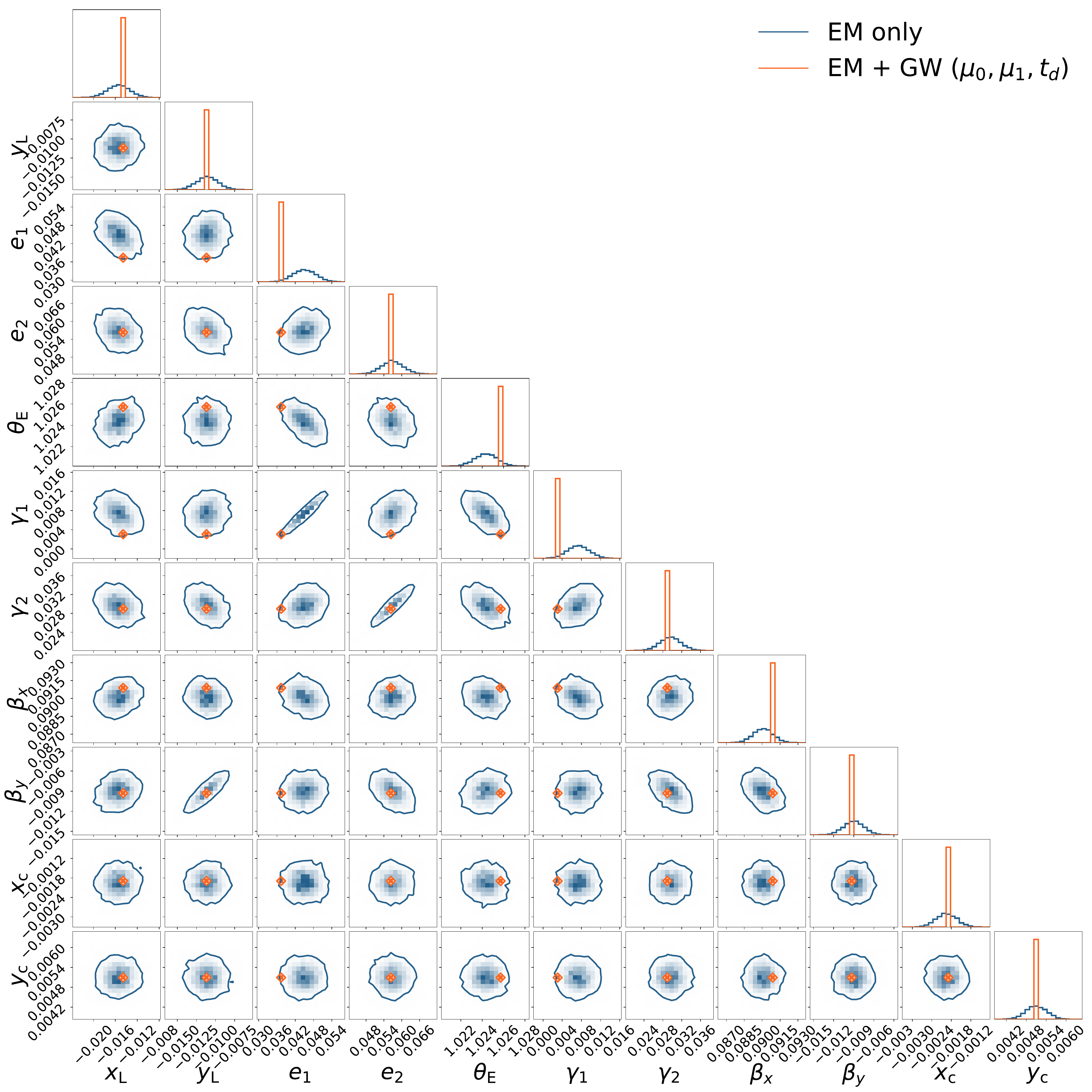}
\caption{EM lens-model posteriors before (blue) and after (orange) including both GW magnifications and the relative time delay, without additional time-delay model-error broadening. The displayed parameters are the lens mass-center coordinates $(x_{\rm L},y_{\rm L})$, ellipticity components $(e_1,e_2)$, Einstein radius $\theta_{\rm E}$, external-shear components $(\gamma_1,\gamma_2)$, source-center coordinates $(\beta_x,\beta_y)$, and lens-galaxy light-center coordinates $(x_c,y_c)$. The latter describe the position of the foreground lens galaxy's light center, distinct from its mass center. Angular quantities are in arcseconds, while ellipticity and shear are dimensionless. The two-dimensional contours enclose $90\%$ of the corresponding joint posterior probability. Although the weight-based ESS is $\Neff=11$, the finite EM sample set cannot resolve the joint posterior.}
\label{fig:lensfull}
\end{figure*}

We next include the GW relative time delay in addition to the two magnifications. Fig.~\ref{fig:lensfull} compares the original EM posterior with the posterior obtained from the full three-dimensional reweighting in $(|\mu_0|,|\mu_1|,\Delta t)$. In contrast to the magnification-only results, the updated distributions become highly concentrated, with extremely narrow one-dimensional features and small, irregular two-dimensional contours. This behavior reflects the limited coverage of the finite EM sample set relative to the very narrow GW timing constraint. For J0252+0039, only 11 of the 54395 EM sample rows receive nonzero weights in the full update, giving the weight-based effective sample size $\Neff=11$. However, these 11 rows are repeated records of a single lens configuration. Thus, the nonzero weight is concentrated on one distinct configuration, and the available samples cannot resolve the joint posterior. The numerical value of the weight-based ESS counts weighted rows and does not account for their repetition or dependence. The concentrated features in Fig.~\ref{fig:lensfull} therefore illustrate inadequate sampling of the joint target, rather than a reliable gain in lens-parameter precision. Any apparent width introduced by histogram binning or smoothing cannot restore the missing posterior coverage. The limitation arises because an independently generated, finite EM posterior provides too few distinct configurations in the region selected by the GW delay. More EM samples with adequate coverage, or sampling directed toward the joint posterior, are needed before the additional timing information can be translated into converged physical constraints.

Overall, the magnification-only results demonstrate that GW observations can provide useful additional constraints on the relative source--lens geometry, even when the improvements in several individual lens parameters remain modest. Incorporating the highly precise GW time delay has the potential to provide further information about the lens model, but its direct application through importance reweighting is limited by the finite coverage of the original EM posterior samples. We examine the impact of introducing an additional time-delay modeling uncertainty in the following subsection.

\subsection{Impact of the time-delay modeling uncertainty}
\label{sec:3.4}

As shown in Sec.~\ref{sec:3.3}, directly incorporating the highly precise GW time-delay constraint into the EM posterior can concentrate the importance weights on very few samples. In addition, the statistical spread of the EM lens-model predictions need not capture all systematic uncertainties in the predicted delay. We therefore investigate the alternative reweighting scheme described in Sec.~\ref{sec:2.3.2}, introducing an additional Gaussian time-delay modeling uncertainty $\sigma_{t,\rm model}$. 

\begin{figure*}[tbp]
\centering
\includegraphics[width=\textwidth]{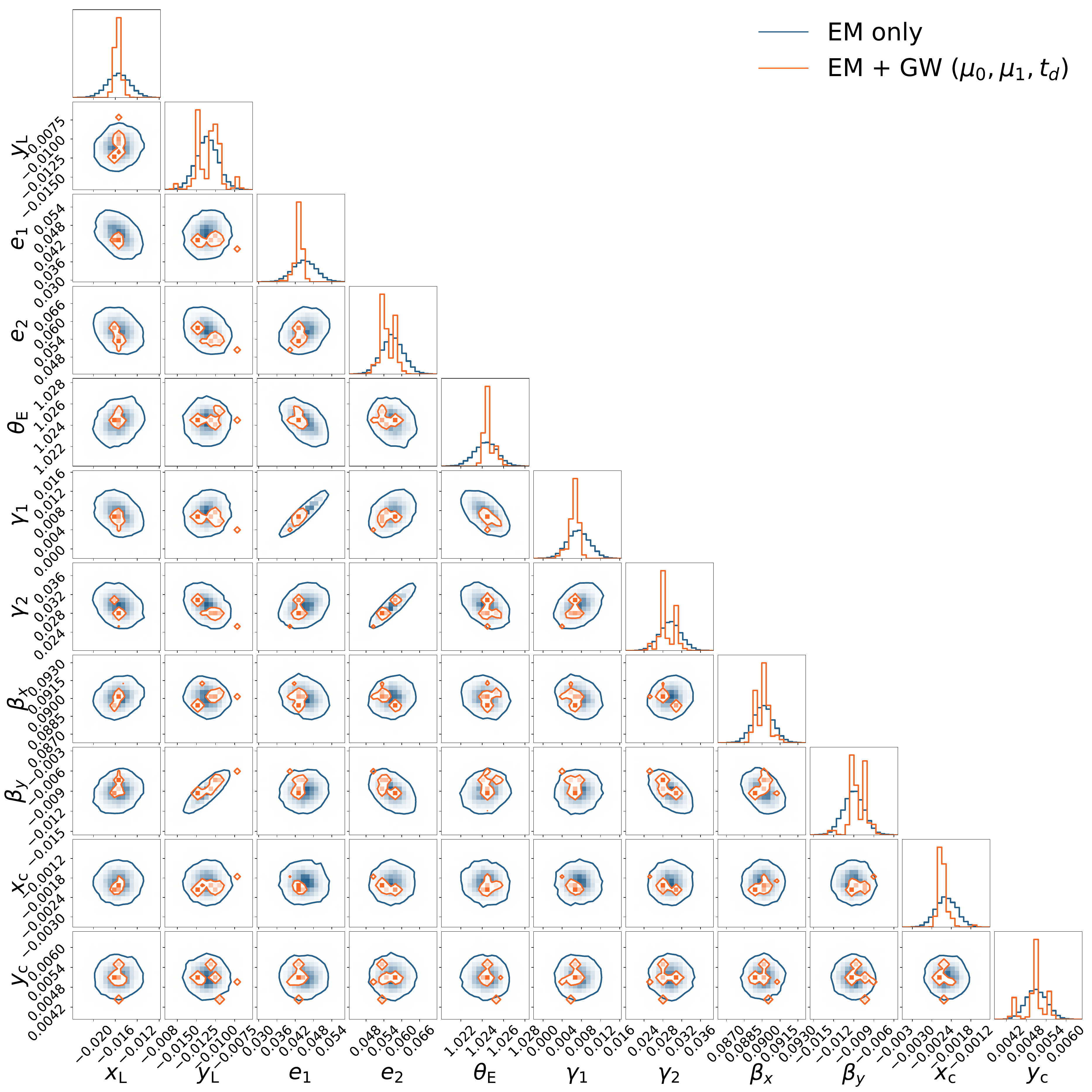}
\caption{EM-only (blue) and GW-updated (orange) lens-model posteriors for J0252+0039 using the two magnifications and the relative delay with $\sigma_{t,\rm model}=10\,\mathrm{s}$. Several updated marginals are substantially narrower, but the effective sample size is only $\Neff\simeq13.9$, so the apparent precision remains sensitive to finite sampling. The displayed parameters are same as Figs.~\ref{fig:lensfull}.}
\label{fig:sigma10}
\end{figure*}

\begin{figure*}[tbp]
\centering
\includegraphics[width=\textwidth]{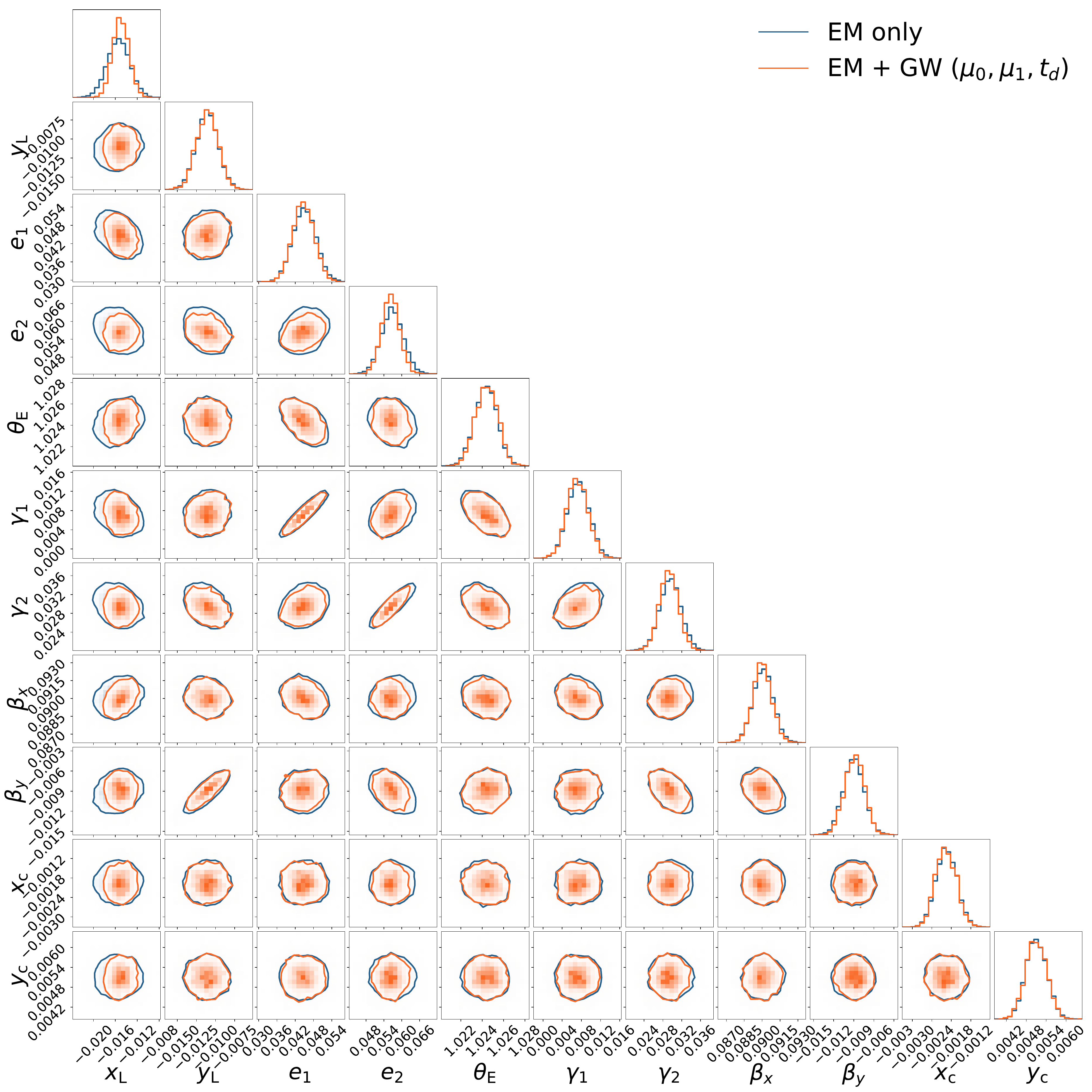}
\caption{EM-only (blue) and GW-updated (orange) lens-model posteriors for J0252+0039 using the two magnifications and the relative time delay, with $\sigma_{t,\rm model}=916374\,\mathrm{s}$, chosen as the integer part of the median time delay predicted by the EM lens model. At this large assumed modeling uncertainty, the timing information is strongly weakened and the update is nearly identical to the magnification-only case, as quantified in Fig.~\ref{fig:three_reduction}. The displayed parameters are same as Figs.~\ref{fig:lensfull}.}
\label{fig:sigma100000}
\end{figure*}

\begin{figure*}[t]
\centering
\includegraphics[width=\textwidth]{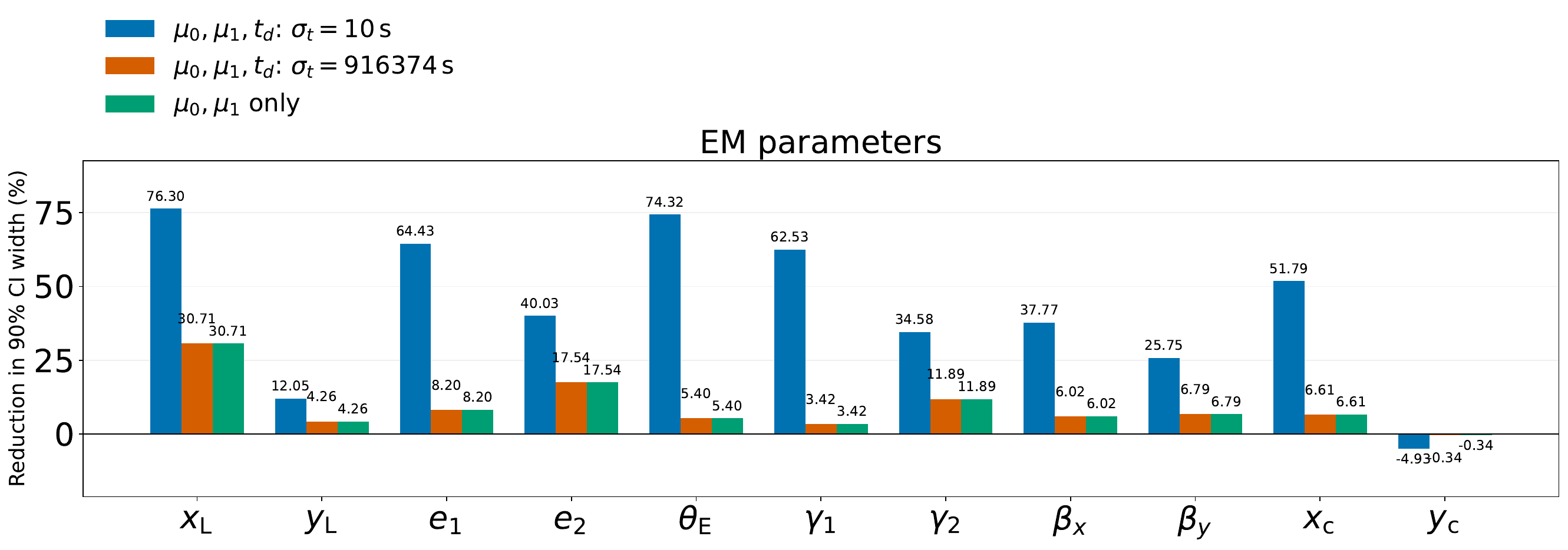}
\caption{Relative reduction in the widths of the 90\% equal-tailed credible intervals of the lens-model and source-position parameters for system J0252+0039 after incorporating GW constraints into the EM analysis. Blue and orange bars include magnification and time-delay constraints with model time-delay uncertainties of $\sigma_t=10\,{\rm s}$ and $916374\,{\rm s}$, respectively; green bars include magnification constraints only. The larger time-delay uncertainty yields reductions nearly identical to the magnification-only case.}
\label{fig:three_reduction}
\end{figure*}

For J0252+0039, we compare $\sigma_{t,\rm model}=10\,\mathrm{s}$ and $\sigma_{t,\rm model}=916374\,\mathrm{s}$, with the latter approximately equal to the reference time delay $\Delta t\simeq916374.25\,\mathrm{s}$. The larger uncertainty serves as a sensitivity test rather than a calibrated estimate of the lens-model error. Figs.~\ref{fig:sigma10} and \ref{fig:sigma100000} show the corresponding lens-parameter posteriors, and Fig.~\ref{fig:three_reduction} compares the relative reductions in the central $90\%$ credible-interval widths, as defined in Sec.~\ref{sec:3.3}. All three cases use the same EM posterior samples as the baseline. The blue bars include both magnifications and the delay with $\sigma_{t,\rm model}=10\,\mathrm{s}$, the orange bars use $916374\,\mathrm{s}$, and the green bars include only the two magnifications. The parameters include the lens mass-center coordinates, ellipticity components, Einstein radius, external-shear components, and source-center coordinates. For $\sigma_{t,\rm model}=10\,\mathrm{s}$, the nominal interval-width reductions are $76.30\%$ for $x_{\rm L}$, $74.32\%$ for $\theta_{\rm E}$, $64.43\%$ for $e_1$, and $62.53\%$ for $\gamma_1$, compared with $30.71\%$, $5.40\%$, $8.20\%$, and $3.42\%$, respectively, in the magnification-only case. The reductions in $\beta_x$ and $\beta_y$ also increase from $6.02\%$ and $6.79\%$ to $37.77\%$ and $25.75\%$. The response is not uniform across parameters: for example, the corresponding reduction in $y_{\rm L}$ is $12.05\%$. These changes are consistent with a narrow timing constraint selecting a restricted set of lens configurations, but the inferred reductions must be assessed together with the weight diagnostics.

In particular, the $10\,\mathrm{s}$ update has a weight-based effective sample size of only $\Neff\simeq13.9$ out of $54395$ EM rows, and the ten largest row weights contain approximately $75.7\%$ of the total probability. The original EM sample set contains only $8313$ distinct nine-parameter lens configurations; the ten highest-weight rows in the $10\,\mathrm{s}$ case represent four distinct configurations. Thus, the broader timing model still selects a region poorly covered by the finite EM sample set. The blue bars in Fig.~\ref{fig:three_reduction} are finite-sample estimates, not demonstrated, converged precision gains. The weight-based ESS does not include corrections for repeated or correlated EM records, and the contour appearance alone cannot establish convergence. By contrast, the orange and green bars are nearly identical for every displayed parameter. The corresponding weight-based effective sample sizes are approximately $16093$ and $16094$, respectively; these values describe weight concentration across rows, not counts of independent configurations. At $\sigma_{t,\rm model}=916374\,\mathrm{s}$, the time-delay factor varies little across the EM configurations carrying appreciable magnification weight, so it adds little information beyond the magnifications. The broader-delay posterior in Fig.~\ref{fig:sigma100000} therefore approaches the magnification-only result. Increasing the time-delay modeling uncertainty reduces weight concentration at the cost of weaker timing constraints, while reliable inference under a narrow time-delay constraint requires improved posterior sampling.

\subsection{Comparison across the ten systems}
\label{sec:3.5}

Table~\ref{tab:ensemble} summarizes the percentage reductions in the central $90\%$ credible-interval widths and the combined optimal SNRs $\rho_c$ for the ten strong-lensing systems. The results illustrate that incorporating EM information improves the GW constraints on both absolute magnifications and the viewing angle, although the degree of improvement varies among systems. The reductions in the credible-interval widths range from $5.0\%$ to $95.3\%$ for $\mu_0^{\rm GW}$, from $7.9\%$ to $68.0\%$ for $\mu_1^{\rm GW}$, and from $7.3\%$ to $52.2\%$ for $\theta_{JN}$.

The different improvements in the two magnifications can be understood from their EM predictions and correlations with the relative time delay. For J0252+0039, the EM lens-model posterior provides a narrower prediction for $\mu_1$ than for $\mu_0$, reflecting their different responses to correlated variations in the lens parameters. The EM constraint on $\mu_1$ is also more informative relative to its original GW posterior. Consequently, magnification-only reweighting reduces the central $90\%$ credible-interval widths of $\mu_0$ and $\mu_1$ by $7.2\%$ and $27.4\%$, respectively. Including the relative time delay further increases these reductions to $26.3\%$ and $68.0\%$, as the joint EM posterior selects magnification combinations consistent with the GW timing measurement. Thus, the unequal improvements arise from the different EM magnification constraints and the additional information provided by their correlations with the time delay. Physically, although the two images originate from the same source and share the same lens mass distribution, they probe different regions of the lens plane and experience different local convergence and shear. As a result, their magnifications respond differently to variations in the lens parameters, leading to different predictive uncertainties in the EM lens-model posterior and, consequently, different improvements after reweighting.

\begin{figure}[tbp]
\centering
\includegraphics[width=\linewidth]{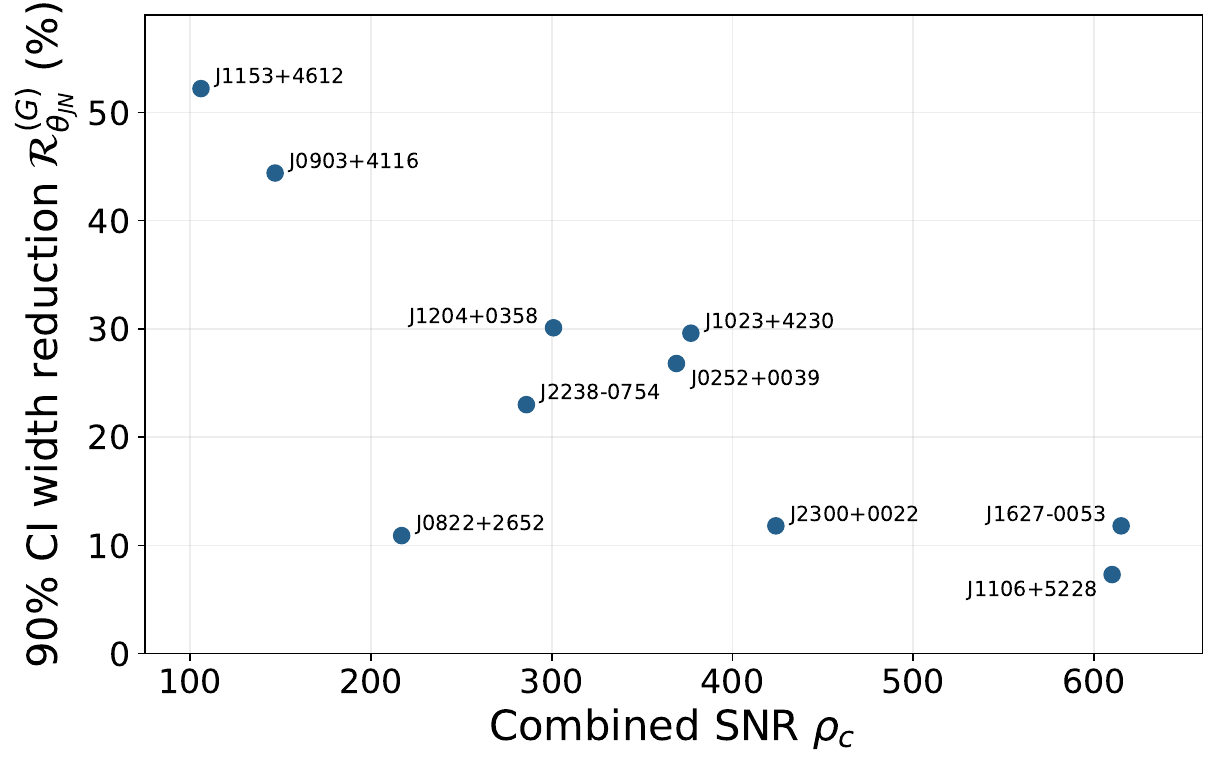}
\caption{Percentage reduction in the central $90\%$ credible-interval width of the viewing angle, $\mathcal R_{\theta_{JN}}^{(G)}$, after incorporating EM lens-model information, as a function of the combined optimal SNR $\rho_c$. Each labeled point represents one of the ten strong-lensing systems. An overall decreasing trend is observed with increasing $\rho_c$, although the degree of improvement varies among systems.}
\label{fig:snr_theta_jn}
\end{figure}

The improvements in the GW parameter constraints also depend on the signal strengths. Across the ten systems, the reductions in the credible-interval widths of $\mu_0^{\rm GW}$ and $\theta_{JN}$ tend to decrease with increasing $\rho_c$, whereas no clear trend is observed for $\mu_1^{\rm GW}$. Figure~\ref{fig:snr_theta_jn} illustrates the relationship between the combined optimal SNR and the improvement in the viewing-angle constraint. An overall negative correlation is observed, indicating that lower-SNR systems tend to benefit more from the additional EM information. For example, J1153+4612, with $\rho_c=106$, exhibits a $52.2\%$ reduction in the viewing-angle interval width, while J1106+5228, with $\rho_c=610$, exhibits a reduction of only $7.3\%$. This trend may reflect the smaller relative contribution of EM information when the original GW constraints are already strong. However, the relationship is not strictly monotonic. For instance, J0822+2652, despite its relatively low combined SNR of $\rho_c=217$, exhibits a reduction of only $10.9\%$. The degree of improvement is therefore not determined by the combined SNR alone, but also depends on the individual image SNRs, the relative constraining power of the EM and GW magnification posteriors, and the correlations between the magnifications and $\theta_{JN}$. Since the systems have different source redshifts, magnifications, and sky positions, these factors vary even though the same intrinsic binary parameters are adopted. 

In the reverse direction, the magnification-only GW-to-EM update reduces the credible-interval widths of the normalized source--lens offset $y$ and the mass--light centroid offset $\Delta r$ by $19.7\%$--$62.3\%$ and $0.1\%$--$54.0\%$, respectively. The improvement in $\Delta r$ varies substantially across systems, reaching $54.0\%$ for J0822+2652 but only $0.1\%$ for J0903+4116. This difference may arise from the different sensitivities of the image magnifications to the mass-center positions and their correlations with the light-center coordinates and other lens parameters. The corresponding changes in $\theta_{\rm E}$ are much smaller, ranging from a $1.1\%$ widening to a $10.8\%$ reduction. The ellipticity amplitude exhibits reductions of $2.1\%$--$40.5\%$, while the changes in the external-shear amplitude range from a $1.5\%$ widening to a $21.5\%$ reduction. Although higher-SNR GW signals can provide more precise magnification measurements, the improvements in the lens parameters show no uniform dependence on $\rho_c$. Instead, these improvements also depend on how the image magnifications respond to variations in the lens parameters and their correlations in the original EM posterior. Since the lens configurations and EM posterior distributions differ among systems, the GW magnification measurements provide different amounts of additional information. In particular, magnification constraints can be more effective for geometric combinations of lens parameters, such as the normalized source--lens offset and the mass--light centroid offset, than for individual parameters affected by degeneracies.

Overall, these results illustrate the complementary roles of GW and EM observations in constraining GW source properties, the relative source--lens geometry, and the spatial alignment between the mass and light distributions of lens galaxies. The efficiency of information transfer depends on both the GW signal strengths and the existing parameter constraints of each system.

\section{Summary and Discussion}
\label{sec:4}

We have investigated joint GW--EM inference using ten real HST-observed strong-lensing systems and simulated pairs of lensed BBH signals. Assuming an identified GW--EM association, a fixed sky position, and a luminosity distance derived from the EM source redshift, we combine existing GW and EM lens-model posteriors through two-way reweighting without refitting the imaging data. Our two-image GW analysis adopts shared source parameters with separate magnifications and arrival times, following previous joint-inference approaches~\citep{Liu2021Joint,Lo2023LensedGW}. While previous studies have explored the identification of lensed GW events and the joint inference of source and lens parameters~\citep{Hannuksela2020Localization,Wempe2024Localization}, our work focuses on quantifying improvements in both GW source and lens-parameter constraints using existing EM lens-model posteriors from real HST observations. We examine both directions of posterior reweighting, with particular attention to the information provided by magnifications and relative time delays.

Across the ten systems, incorporating EM lens-model information reduces the central $90\%$ credible-interval widths of the two GW magnifications by $5.0$--$95.3\%$ and $7.9$--$68.0\%$, respectively, and that of the viewing angle by $7.3$--$52.2\%$. The mass parameters and relative time delay are generally less affected. Conversely, GW magnification measurements reduce the interval widths of the normalized source--lens offset, mass--light centroid offset, and ellipticity amplitude by $19.7$--$62.3\%$, $0.1$--$54.0\%$, and $2.1$--$40.5\%$, respectively. The Einstein radius and external-shear amplitude show reductions of up to $10.8\%$ and $21.5\%$, with slight broadening in some systems. These results demonstrate that GW magnification measurements can improve constraints on combinations of lens parameters, particularly those describing the relative source--lens geometry and the spatial alignment between the mass and light distributions of the lens galaxy, even when individual parameters show modest improvements. The viewing-angle improvement arises from its correlation with the magnifications, since the luminosity distance and sky position are fixed in the GW analysis.

A major limitation arises when incorporating the highly precise GW relative time delay into the EM posterior. For J0252+0039, the full reweighting assigns nonzero weights to only 11 repeated records corresponding to a single lens configuration, indicating insufficient sampling of the joint posterior. Introducing a $10\,\mathrm{s}$ time-delay modeling uncertainty increases the weight-based effective sample size to approximately $13.9$, but the resulting posterior remains dominated by a few configurations. Therefore, the apparent improvements in lens-parameter precision cannot be considered reliable without adequate posterior sampling. Increasing the modeling uncertainty to $916374\,\mathrm{s}$ reduces weight concentration but produces results nearly identical to magnification-only reweighting. This illustrates the trade-off between the additional timing information and the stability of importance reweighting. Reliable inference requires improved coverage of the joint posterior and physically motivated time-delay modeling uncertainties~\citep{ChatterjeeDiaconis2018,Vehtari2024PSIS,Birrer2025Multimessenger}.

Several assumptions also limit the interpretation of our results. We assume that the BBH merger occurs at the reconstructed light center of the background galaxy, where the lensing magnifications and time delays are calculated. For the GW injection and parameter estimation, we approximate the sky positions of the lensed images using the foreground lens coordinates (RA and DEC), neglecting the small angular separations between the images. The inferred source offset therefore describes the relative position of the background-galaxy light center and the lens mass center, rather than an independent localization of the BBH within its host galaxy. An offset between the merger and the source light center could modify the image magnifications and time delays. Future analyses should account for the uncertainty in the merger position and its impact on the lensing observables.

Our lens constraints are conditional on the adopted singular isothermal ellipsoid (SIE) model with external shear. Since the model does not include a free mass-sheet or external-convergence parameter, the resulting improvements do not demonstrate that the mass-sheet degeneracy has been broken~\citep{SchneiderSluse2013,SchneiderSluse2014,Poon2025Degeneracies}. More flexible mass models may change the inferred constraints and should be investigated in future work.

Microlensing and image multiplicity introduce additional considerations. Compact objects in the foreground lens galaxy can produce microlensing effects that modify the GW waveform through frequency-dependent magnification and phase shifts, potentially biasing the inferred magnifications if only macrolensing is modeled~\citep{Diego2019,Cheung2021,Mishra2021}. Furthermore, although our analysis considers pairs of lensed GW signals, some strong-lensing systems can produce four images. In such cases, the selected image pair must be consistently identified across lens-model realizations. If more than two GW images are detected, a joint analysis of all available images could provide additional magnification and time-delay constraints, potentially improving the inference of the source and lens parameters. Extending the present framework to multiple images while accounting for microlensing effects is therefore an important direction for future work.

Finally, our simulations adopt identical binary source parameters across the ten systems and fix several parameters during GW inference. More realistic populations, different noise realizations, and broader parameter sampling are needed to assess the generality of the reported improvements. The GW--EM association is assumed to be known, and our results do not quantify the probability of a correct association or the effects of observational selection~\citep{LiuLiao2026,Mandel2019Selection,Vitale2022Selection}. The ten systems should therefore be regarded as representative case studies rather than a population forecast.

Overall, our framework provides a practical approach to joint GW--EM parameter estimation using existing EM lens-model posteriors without refitting the imaging data. The combined observations improve constraints on GW source orientation and source--lens geometry, while fully exploiting precise GW time-delay measurements requires improved posterior sampling and realistic lens-model uncertainties.

\begin{acknowledgments}
We thank Guopeng Li for useful discussion.
\end{acknowledgments}


%

\end{document}